\documentclass[aps,prd,twocolumn,amsmath,showpacs,amssymb,superscriptaddress,nofootinbib,longbibliography,eqsecnum]{revtex4-1}
\usepackage{graphicx}
\usepackage{bm}
\usepackage{amssymb,amsmath}
\usepackage{mathrsfs}
\usepackage{latexsym}
\usepackage{color}
\usepackage[normalem]{ulem} 
\usepackage{dcolumn}
\usepackage[colorlinks=true,citecolor=blue,urlcolor=blue]{hyperref}
\usepackage[usenames,dvipsnames]{xcolor}

\allowdisplaybreaks

\newcommand{\Caltech}{\affiliation{Theoretical Astrophysics 350-17,
    California Institute of Technology, Pasadena, CA 91125}}
\newcommand{\CITA}{\affiliation{CITA, 60 Saint George Street, Toronto, ON, M5S 3H8}}

\newcommand{\ba}{\begin{align}}
\newcommand{\ea}{\end{align}}
\newcommand{\bma}{\begin{pmatrix}}
\newcommand{\ema}{\end{pmatrix}}

\newcommand{\aaron}[1]{\textcolor{blue}{#1}}

\begin{document}

\title{Damped and zero-damped quasinormal modes of charged, nearly extremal black holes}

\author{Aaron Zimmerman}
\CITA
\author{Zachary Mark}
\Caltech
\date{\today}

\begin{abstract}
Despite recent progress, the complete understanding of the perturbations of charged, rotating black holes as described by the Kerr-Newman metric remains an open and fundamental problem in relativity.
In this study, we explore the existence of families of quasinormal modes of Kerr-Newman black holes whose decay rates limit to zero at extremality, called zero-damped modes in past studies.  
We review the nearly extremal and WKB approximation methods for spin-weighted scalar fields (governed by the Dudley-Finley equation) and give an accounting of the regimes where scalar zero-damped and damped modes exist.
Using Leaver's continued fraction method, we verify that these approximations give accurate predictions for the frequencies in their regimes of validity.
In the nonrotating limit, we argue that gravito-electromagnetic perturbations of nearly extremal Reissner-Nordstr\"{o}m black holes have zero-damped modes in addition to the well-known spectrum of damped modes.
We provide an analytic formula for the frequencies of these modes, verify their existence using a numerical search, and demonstrate the accuracy of our formula. 
These results, along with recent numerical studies, point to the existence of a simple universal equation for the frequencies of zero-damped gravito-electromagnetic modes of Kerr-Newman black holes, whose precise form remains an open question.
\end{abstract}


\maketitle
 
\section{Introduction}

The Kerr-Newman (KN) black hole~\cite{Newman:1965my,Adamo:2014baa,Chrusciel:2012jk} is the most general four-dimensional black hole solution to the electro-vacuum Einstein field equations, provided that the unphysical magnetic and NUT charges are set to zero. 
While astrophysical black holes cannot maintain significant charge~\cite{Gibbons1974,Blandford:1977ds}, charged black holes remain fundamental objects of study in gravitational and quantum theories. 
KN black holes are the simplest charged, rotating objects allowed by relativity, and so provide a natural arena to study the interplay of electromagnetism and gravity. 
However, perturbations of these black holes have until recently been poorly understood, even many years after their discovery.

As in other black hole solutions, perturbed KN black holes possess a spectrum of decaying, resonant oscillations. 
These quasinormal modes (QNMs)~\cite{Kokkotas1999,Berti2009} are excited by transient sources, and they decay as energy flows into the black hole horizon and outward to asymptotic infinity. 
In simpler black hole solutions, such as the rotating Kerr black hole~\cite{Kerr:1963ud,Teukolsky:2014vca}, the quasinormal modes can be understood as the eigensolutions to systems of ordinary differential equations, with the QNM frequencies given by the eigenvalues. 
The study of QNMs is an essential topic in understanding the structure of black hole spacetimes. 
QNMs play a role in gravitational wave astrophysics (e.g.~\cite{Dreyer2004,Berti:2005ys,Barausse:2014tra}) where they make up the ``ringdown'' following the birth of a black hole or the merger of two black holes; have connections to quantum field theories through the AdS/CFT correspondence~\cite{Son:2007vk}; and are potentially linked to quantum mechanical excitations of black holes (see the extensive references in~\cite{Berti2009}).

While much is known about the spectrum of Kerr black holes and the nonrotating, charged Reissner-Nordstr{\"o}m (RN) black holes, the investigation of the QNMs of KN black holes has proven difficult outside of the scalar case discussed below.
Gravitational and electromagnetic perturbations of the Kerr black hole can be tackled by studying spin-weighted scalar fields propagating on the black hole background.
These scalars obey a master equation~\cite{Teukolsky1973}, which separates into coupled ordinary differential equations.
When these same methods are applied to applied to the KN black hole, the result is a system of equations describing coupled gravitational and electromagnetic perturbations~\cite{ChandraBook,Dias:2015wqa}. While these equations can still be expanded in frequency and azimuthal harmonics due to the symmetries of the spacetime, they have not been separated in the remaining coordinates. 
In contrast, for the spherically symmetric RN black hole, separation is possible and equations for coupled ``gravito-electromagnetic'' (GEM) perturbations can be derived~\cite{Moncrief:1974ng, Moncrief:1974gw, Zerilli:1974ai}. Using these equations, the QNMs of RN have been extensively studied~\cite{Berti2009}.

The difficulties in analyzing the perturbations of KN black holes led to the exploration of simpler wave equations on the KN background, in the hope that they might provide a reasonable approximation to the full problem. 
In particular, Kokkotas and Berti studied the QNMs of the Dudley-Finley (DF) equation on the KN backgrounds~\cite{Kokkotas:1993ef,Berti:2005eb}. 
The DF equation~\cite{Dudley:1977zz,Dudley:1978vd} describes the propagation of spin-weighted test fields in various spacetimes. 
Of particular interest is type D spacetimes such as KN, where the DF equation separates. 
In this case, the DF equation can be reduced to a coupled eigenvalue problem for the QNMs, just as for the perturbations of the Kerr black hole.
In the spin-zero case, $s=0$, the DF equation reduces to the wave equation for a massless, uncharged scalar field, $\nabla^\mu \nabla_\mu \psi = 0$. Thus the analysis of the $s = 0$ DF equation yields the true scalar QNMs of the KN black hole. 
For $s\neq 0$, the QNMs can only be an approximation (and possibly a poor one) to the QNMs of the gravitational and electromagnetic perturbations of KN.
The DF equations assume that each test field is treated independently\footnote{For example, one can show using the methods in~\cite{Teukolsky1973} that the DF equation is obeyed by a second fictitious set of Maxwell fields propagating on the KN background, which have no leading order background contributions.}, which does not correctly capture the coupling between the electromagnetic and gravitational perturbations of KN.
Although there has been some confusion in the literature on this point, there is no reason {\it a priori} to expect that the DF equation provides anything more than a qualitative description of the QNMs of KN.

Recently, a great deal of progress has been made in understanding the true GEM modes of KN black holes. 
New approximation techniques have allowed for the investigation of KN black holes which deviate from Kerr and RN black holes by small amounts. 
Slowly rotating, charged black holes were treated using using a matrix-valued continued fraction method~\cite{Pani:2013ija,Pani:2013wsa}. 
Following this, Mark {\it et al.}~\cite{Mark:2014aja} tackled the case of a weakly charged KN black hole, using an eigenvalue perturbation method adapted for quasinormal modes~\cite{Yang:2014tla, Zimmerman:2014aha}.
Most excitingly, challenging numerical studies have allowed the exploration of the QNMs for the full range of angular momentum and charge parameters of KN black holes for the first time~\cite{Zilhao:2014wqa, Dias:2015wqa}. 
The work of Dias, Godazgar, and Santos~\cite{Dias:2015wqa} is especially noteworthy, presenting a complete scan of the $(a,\,Q)$ parameter space of KN black holes, for the lowest overtone and $l\leq 3$.
While this comprehensive information is now in principle available, an analytic understanding of these QNMs remains an important goal.

One region of parameter space which is of special interest and which may be amenable to analytic techniques is the nearly extremal KN (NEKN) black holes. Kerr-Newman black holes have an extremal combination of charge and angular momentum, which causes their inner and outer horizons to coalesce, and where the surface gravity of the black hole vanishes.
Beyond these extremal combinations of charge and spin the singularity within the black hole is naked to asymptotic infinity, destroying any notion of causality. 
As black holes enter the nearly extremal regime a new approximation scheme  becomes available, in terms of the small distance from extremality. 
For the Kerr spacetime, the Teukolsky equation simplifies in this limit and approximate formulas for the QNM frequencies are known.
These formulas describe weakly damped QNMs with the real part of the frequency $\omega$ given by $m \Omega_H$ and the imaginary part proportional to $\sqrt{\epsilon}$~\cite{TeukolskyPress1974,Detweiler1980,Andersson2000,Glampedakis2001,Hod2008a,Yang:2012he,Yang:2012pj,Yang:2013uba}, where $\epsilon  = 1- a/M \ll 1$ is the small expansion 
parameter. 
These modes have been well studied, and were called the zero-damped modes (ZDMs) in~\cite{Yang:2012pj,Yang:2013uba} to distinguish them from a second family of modes, the damped modes (DMs) whose decay remains nonzero in the limit $\epsilon \to 0$. 
The ZDMs can be analyzed in the nearly extremal limit using a matched asymptotic expansion of the radial equation, and appear to be related to the mathematical horizon instability of extremal black holes~\cite{Aretakis:2012ei,Aretakis:2013dpa,Reiris:2013efa}.
Meanwhile, the DMs must be treated using a different method, either through numerical exploration or the use of a WKB analysis in Kerr~\cite{Yang:2012he}.

Mark {\it et al.}~\cite{Mark:2014aja} were able to deal with the case of rapidly rotatng, weakly charged KN black holes. The results of that study hint at the existence of ZDMs in these spacetimes, and surprisingly also show that the DF equation alone provides the correct small-charge perturbation to the ZDM frequencies of Kerr.
In addition, the recent numerical studies of KN QNMs~\cite{Zilhao:2014wqa, Dias:2015wqa} provide strong evidence that for NEKN black holes, the QNM frequencies are described by an equation like that obeyed by the ZDMs of Kerr. 
With these results, one might hope that all the modes of NEKN might be described by a simple equation.

These facts motivate the exploration of both the DF equation and the full GEM equations of KN in the nearly extremal limit.
The goal of this study is more modest; we give an accounting of those cases where the full problem simplifies.
In Sec.~\ref{sec:DF} we study the DF equation for NEKN black holes. We identify ZDMs which exist for any value of $a$, including the nonrotating case of the nearly extremal RN (NERN) black hole, by combining analytic approximations and numerical mode searches.
A WKB analysis valid for scalar fields on KN~\cite{HodEikonal2012,Zhao:2015pqa} identifies where modes with nonzero decay exist in the parameter space of extremal KN black holes, implying that in these cases the QNM spectrum bifurcates into two branches as the black hole approaches extremality at fixed angular momentum. 
This bifurcation occurs for certain QNM families of Kerr, as discussed in~\cite{Yang:2012pj,Yang:2013uba}.
We discuss this WKB analysis in Sec.~\ref{sec:WKBanalysis}.
In Sec~\ref{sec:DFNumerics} we use a numerical search to determine the accuracy of the nearly extremal and WKB approximations, and we also find separated families of DMs and ZDMs.
Note that the frequency formula we derive here for scalar modes of NEKN black holes was presented without derivation in recent note~\cite{Hod:2014uqa}.

Motivated by these results for scalar fields, in Sec.~\ref{sec:RN} we show that ZDMs also exist in the case of the GEM modes of NERN, using an analytic approximation and matching ansatz. We confirm the existence of these modes with an explicit numerical search.
These modes are purely decaying with a small decay rate, and they appear to have been overlooked despite a long history of study of the QNMs of RN. 
Such modes give additional support to the conjecture that ZDMs exist for all nearly extremal black holes, but they also demonstrate the failure of the DF equation to accurately capture the GEM frequencies in this limit.

Finally, we discuss prospects for the open problem of the coupled GEM perturbations of NEKN in Sec.~\ref{sec:Conclusions}.
In this paper, we do not attempt to discuss the topic of charged and massive fields on KN backgrounds where superradiant instabilities are found to arise (see e.g.~\cite{Brito:2015oca} for a recent review).

Throughout this paper we use geometric units so that charge and mass have units of length, setting $G = c =1$. We provide a reference for the definitions of some of the important variables used in Table~\ref{tab:VarList}.

\begin{table}
\renewcommand{\arraystretch}{2}
\caption{
\label{tab:VarList}
List of relevant definitions
}
\begin{ruledtabular}
\begin{tabular}{@{} l c l @{}}
$r_\pm$ & $\displaystyle M \pm \sqrt{M^2 - a^2 -Q^2}$ & Horizon positions \\
$\sigma$ & $\displaystyle \frac{r_+ - r_-}{r_+}$ & Near-extremal parameter \\
$k$ & $\omega - m \Omega_H$ &  Corotating frequency \\
$\hat{\omega}$ & $\omega r_+$ & Dimensionless frequency \\
$ \Omega_H$ & $\displaystyle \frac{a}{r_+^2 +a^2}$ & Horizon frequency \\
$\kappa$ & $\displaystyle  \frac{\sigma r_+}{2(r_+^2 +a^2)}$ & Surface gravity \\
$\mathcal J^2$ & $(m \Omega_H)^2(6M^2 +a^2) - A$ & WKB indicator of DMs \\
$\delta^2$ & $\displaystyle 4 \hat{\omega}^2- (s+1/2)^2 - \lambda$ & DF matching parameter \\
$\varpi$ & $\displaystyle \frac{k}{\kappa}  - i s $ & DF matching parameter \\ 
$\zeta$ & $2m r_+ \Omega_H - i s$ & DF matching parameter \\ 
$x$ & $\displaystyle \frac{r-r_+}{r_+} $ & Radial parameter \\
$\displaystyle{\genfrac{}{}{0pt}{}{\omega_R,} {\omega_I}}$ & $\omega = \omega_R + i \omega_I$ & Frequency components
\end{tabular}
\end{ruledtabular}
\end{table}

\section{The Dudley-Finely equation for nearly extremal Kerr-Newman black holes}
\label{sec:DF}

In this section we discuss the QNMs of the Dudley-Finley equation for  nearly extremal Kerr-Newman black holes.

Due to the similarity between the Teukolsky equation and the DF equation, Leaver's continued fraction method~\cite{Leaver1986}, used to accurately compute the QNMs of Kerr, extends easily to the DF equation in the KN background~\cite{Berti:2005eb}.
This method can in principle provide accurate QNM frequencies for any spin angular momentum, charge, and harmonic.
Since our interest is to develop further analytic understanding of the QNM frequencies in the limit of nearly extremal black holes, we use the nearly extremal and WKB approximations to explore the ZDM and DM frequencies. 
We then use Leaver's method to confirm these approximations, and to measure their error.

\subsection{Kerr-Newman black holes}

Kerr-Newman black holes are parametrized by their mass $M$, specific angular momentum $a$, and charge $Q$ when magnetic and NUT charges are neglected. One convenient way to represent the metric for the KN spacetime in Boyer-Lindquist coordinates~\cite{BoyerLindquist1967} is obtained by writing the line element of Kerr in terms of the second degree polynomial $\Delta = r^2 - 2 M r+a^2$. The roots of this polynomial are the inner ($r_-$) and outer $(r_+)$ horizons of Kerr, $\Delta = (r-r_+)(r-r_-)$. The Kerr-Newman metric follows from using the appropriate definition of $\Delta$ when charge is included. This form of the line element is~\cite{MTW}
\begin{align}
ds^2  &=  - \frac{\Delta}{\rho^2 } \left(dt - a \sin^2\theta d \phi \right)^2 + \frac{\rho^2}{\Delta} dr^2 + \rho^2 d\theta^2 \notag \\
& \, + \frac{\sin^2 \theta}{\rho^2} \left[a dt - (r^2+a^2)d\phi\right]^2
\,, \\
\Delta  &=  r^2 - 2 M r +a^2 +Q^2 \,, \\
\rho^2  &=  r^2 +a^2 \cos^2 \theta\,.
\end{align}
The corresponding vector potential takes the form 
\begin{align}
A_\mu dx^\mu =\frac{Qr}{\rho^2}\left(dt -a\sin^2\theta d\phi \right)\,.\end{align}
The outer and inner horizons are located at
\begin{align}
r_\pm =  M \pm \sqrt{M^2 - a^2 - Q^2} \,.
\end{align}
In the nearly extremal limit, where the inner and outer horizons approach each other, we define the small 
parameter\footnote{In~\cite{Yang:2013uba} the small parameter $\epsilon$ is used, while older studies use $\sigma$ as the small parameter. 
Using $\sigma$, the analysis of nearly extremal Kerr carries over naturally to KN black holes.}
$\sigma \ll1$ as

\begin{align}
\sigma = \frac{r_+ - r_-}{r_+} \approx 2 \sqrt{1 - a^2/M^2 - Q^2/M^2} \,.
\end{align}
It is also useful to recall the expression for the surface gravity $\kappa$ of the KN black hole~\cite{Wald1984}
\begin{align}
\kappa = \frac{r_+ - r_-}{2(r_+^2 +a^2)} = \frac{\sigma r_+}{2(r_+^2 +a^2)} \,.
\end{align}
We see that $r_+ \kappa \ll 1$ in the nearly extremal limit.

\subsection{The Dudley-Finley equation}

We turn to the analysis of the DF equations in the KN spacetime. 
These equations and their analysis closely parallels the treatment of scalar, electromagnetic, and gravitational perturbations of the Kerr spacetime by \aaron{using} spin-weight $s$ scalars ${}_s \psi$, which obey a separable master equation~\cite{Teukolsky1973}.
Just as in Kerr, we expand the spin-weighted scalars ${}_s \psi$ in frequency and azimuthal harmonics as
\begin{align}
{}_s \psi & = \sum_{lm} \int d\omega \, e^{-i(\omega t - m \phi)} {}_s R_{lm\omega}(r) {}_sS_{lm\omega}(\theta) \,.
\end{align}
With this expansion, the DF wave equations in the KN spacetime separate~\cite{Kokkotas:1993ef} and are nearly identical to the corresponding equations in Kerr.

The angular functions $ {}_sS_{lm\omega}(\theta)$ are spin-weighted spheroidal harmonics and obey the angular Teukolsky equation~\cite{Teukolsky1973,Fackerell1977,Berti2009}, 
\begin{align}
\label{eq:TeukS}
\csc \theta & \frac{d}{d\theta} \left( \sin \theta  \frac{dS}{d\theta}\right) + V_\theta S = 0 \,, \\
V_\theta = & \, a^2 \omega^2 \cos^2 \theta - m^2 \csc^2 \theta - 2 a \omega s \cos \theta \notag  \\ &
- 2 m s \cot \theta \csc \theta -s^2 \cot^2 \theta + s + {}_s A_{lm \omega} \,.
\end{align} 
Here the angular separation constants for each harmonic are denoted ${}_s A_{lm\omega}$.
In the limit of a Schwarzschild black hole ($a \to 0$) they simplify, $A \to l(l+1) - s(s+1)$.

The radial functions ${}_s R_{lm\omega}(r)$ obey a second order differential equation. In the source-free case it is
\begin{align}
\label{eq:TeukR}
\Delta^{-s} & \frac{d}{dr} \Delta^{s+1} \frac{dR}{dr} + V_r R = 0 \,, \\
\label{eq:PotR}
V_r & =  \frac{K^2 + i s K \partial_r \Delta }{\Delta} - 2 i s \partial_r K - {}_s\lambda_{lm\omega} \,, \\
K & = - \omega(r^2+a^2) + am \,, \\
{}_s \lambda_{lm \omega} & = {}_sA_{lm\omega} - 2 a m \omega +a^2 \omega^2\,.
\end{align}
Here and elsewhere we suppress spin-weight and harmonic indices where there is no danger of confusion.

It is useful to define a tortoise coordinate $r_*$ and rewrite the radial equation in terms of a different function ${}_s u_{lm\omega}$. These are defined as follows:
\begin{align}
\frac{dr_*} {dr}& = \frac{r^2 +a ^2} {\Delta} \,, &   u & = \Delta^{s/2} \sqrt{r^2 +a^2} R \,.
\end{align}
With these substitutions, the radial equation~\eqref{eq:TeukR} becomes
\begin{align}
\label{eq:Teuku}
\frac{d^2 u}{dr_*^2}& + V_u u = 0 \,,\\
V_u & = \frac{K^2 + 2 i s K (r- M) + \Delta(4 i s \omega r - \lambda)}{(r^2+a^2)^2} - G^2 -\frac{dG}{dr_*} \,,\\
G & = \frac{\Delta r}{(r^2+a^2)^2} + \frac{s \partial_r \Delta}{2(r^2+a^2)} \,.
\end{align}
Equation~\eqref{eq:Teuku} makes it apparent that the asymptotic solution as $r_* \to - \infty$ (as $r\to r_+$) is the same as in Kerr but with a change in the definition of $k$ and $\Delta$,
\begin{align}
\label{eq:Ingoingu}
u &\sim \exp\left[\pm \frac{s}{2} \frac{\sigma r_+}{r_+^2 +a^2} r_* \right]e^{\pm i k r_*}  = \Delta^{\pm s/2} e^{\pm i k r_*}
 \,, \\
\label{eq:Defk}
k& = \omega - m \frac{a}{r_+^2 +a^2} = \omega - m \Omega_H \,,
\end{align}
where we have identified the horizon frequency $\Omega_H$ in the final expression. 
The asymptotic solution $u \sim \Delta^{-s/2} e^{-ikr_*}$ corresponds to waves traveling into the horizon \cite{Teukolsky1973}.
The second asymptotic solution corresponds to waves which are directed out of the horizon, and we discard this unphysical solution.
Meanwhile, the outer asymptotic solutions remain unchanged from Kerr to KN,
\begin{align}
\label{eq:Outgoingu}
u \sim r^{\pm s} e^{\mp i \omega r_*} && r \to \infty \,.
\end{align}
These solutions correspond to ingoing waves for $u\sim e^{+i\omega r_*}$ and outgoing waves for $u\sim e^{-i\omega r_*}$.
The QNMs of the DF equation can be found by solving the radial equation~\eqref{eq:Teuku} with an outgoing boundary condition; only certain frequencies allow for solutions which also obey the boundary condition at the horizon.

Both the matched asymptotic expansion and the WKB analysis of Kerr carry directly over to the case of the DF equation in KN. 
In the remainder of this section, we discuss these results, which demonstrate the bifurcation of the scalar spectrum of the NEKN black hole into ZDMs and DMs.
These results also set the stage for an understanding of the existence of both ZDMs and DMs for the coupled GEM perturbations of NERN black holes as discussed in Sec.~\ref{sec:RN}. 
For NEKN black holes, either the spin parameter $a$ or the charge $Q$ can be eliminated in favor of $\sigma$. 
We choose to retain an explicit dependence on $a$ in our equations.

\subsection{Matching analysis of the Dudley-Finley equation in nearly extremal Kerr Newman}
\label{sec:Matching}

To investigate the nearly extremal modes, we must use the technique of matched asymptotic expansions\footnote{A regular perturbation analysis where the wave function is a power series in $\sigma$ does not work because, in the language of matched asymptotic expansions, there is a boundary layer at the horizon.}.
Our analysis closely follows the analysis of the Kerr case in Yang {\it et al.}~\cite{Yang:2013uba} and the references therein. 
For this we define a new coordinate variable $x$ and dimensionless frequency $\hat \omega$ via
\begin{align}
x & = \frac{r - r_+}{r_+} \,,  & \hat \omega &= \omega r_+ \,.
\end{align}
The method splits the region exterior to the black hole into an outer, far-field region $x \gg \sigma$, and an inner, near-horizon region $x \ll 1$. 
The method only works for ZDMs; that is, we assume from the beginning that the frequency has the form $\omega = m \Omega_H + O(\sigma)$. 
First we look at radial equation in the outer region.

\subsubsection{The outer solution}

We rewrite the Teukolsky equation for $R$, Eq.~\eqref{eq:TeukR}, in terms of $x,\, \hat \omega$ and substitute $k$ from Eq.~ \eqref{eq:Defk}.
Using the approximation $\sigma\ll x$, we arrive at the same outer solution as in Kerr,
\begin{align}
\label{eq:TeukFar}
x^2 R'' + 2 (s+1) x R' + \left[\hat \omega^2 (x+2)^2 + 2 i s \hat \omega x - \lambda \right] R = 0 \,.
\end{align}
We define $\delta$ through
\begin{align}
{}_s\delta_{lm\omega}^2 = 4 \hat \omega^2 - (s+1/2)^2 - {}_s\lambda_{lm\omega} \,,
\end{align}
and we find the solution to Eq.~\eqref{eq:TeukFar} in terms of confluent hypergeometric functions ${}_1F_1$~\cite{nist},
\begin{align}
\label{eq:OuterSln}
R =  A &e^{-i \hat \omega x} x^{ -1/2 - s + i \delta} \notag \\
& \times {}_1F_1(1/2-s + i \delta + 2i \hat \omega, 1 + 2 i \delta, 2 i \hat \omega x) \notag \\
& + B (\delta \to - \delta) \,,
\end{align}
where $(\delta \to - \delta)$ indicates that the same functions are repeated with the sign of $\delta$ reversed.
In NEKN $\delta$ no longer takes the explicit form it has in Kerr, $\delta_{\rm K}^2 = 7m^2 / 4 -(s+1/2)^2 - A$, because while the frequency still becomes nearly proportional to the horizon frequency, $\omega \to m \Omega_H$, here the horizon frequency is a varying function of $a$.

The outgoing wave condition for QNM frequencies imposes one constraint on $A$ and $B$. This condition is derived by expanding the ${}_1F_1$ functions at large $x$ into ingoing and outgoing parts, and requiring a cancellation of the ingoing waves.
We provide the condition in Appendix~\ref{sec:MatchingApp}, since there are a few minor sign errors\footnote{Specifically, the factors of $s$ in Eq.~(3.9) have the wrong signs, which is countered by an identical error in Eq.~(3.16).} in~\cite{Yang:2013uba}.
In order to get a second condition and derive the QNMs frequencies, we turn to the inner solution in the near-horizon region.

\subsubsection{The inner solution}

The next step is to assume that both $x \ll1$ and $\sigma \ll 1$ but make no assumptions about their relative size. We use Eq.~\eqref{eq:Teuku} as our starting point, make the substitutions for $x$ and $\sigma$, and further note that the quantity $M k / \sigma$ can be order unity. Expanding the potential $V_u$ to second order in the small quantities while keeping factors of $k/\sigma$ intact gives our desired potential. 
A second coordinate transform brings the radial equation into a simpler, more tractable form.

The second transformation is performed by noting that in the near-horizon region, we can approximate $r_*$ by
\begin{align}
r_* \approx \frac{1}{2\kappa}  \ln \left(\frac{x}{x+\sigma} \right) = \frac{r_+^2 +a^2}{\sigma r_+} \ln \left(\frac{x}{x+\sigma} \right) \,.
\end{align}
We then define $y$ through
\begin{align}
y & = e^{2 \kappa r_* } \approx \frac{x}{x+\sigma} \,, &
\frac{d}{dr_*} & = 2 \kappa y \frac{d}{dy} \,,
\end{align}
which in the Kerr spacetime limits to $y = \exp(\sqrt{2 \epsilon} r_*)$. 
After transforming to the $y$ coordinate the second derivatives in the radial equation become
\begin{align}
\frac{d^2 u}{dr_*^2} & = \left( 2 \kappa \right)^2 \left(y^2 \frac{d^2 u}{dy^2} + y \frac{d u}{dy} \right) \,.
\end{align}
Substituting $y$ for $x$ in the expansion of $V_u$ and dividing out the prefactor gives us our differential equation for $u$. 
It is useful to make the following 
definitions\footnote{When comparing these to the results in~\cite{Yang:2013uba} it is useful to note that  $\varpi = \sqrt{2} \bar \omega$ and $\zeta = \bar m$, in the notation of that paper.},
\begin{align}
\varpi & = \frac{k}{\kappa}  - i s \,, &  \zeta = 2 m r_+  \Omega_H - i s \,.
\end{align}
Then we have
\begin{align}
\label{eq:NHdiffeq}
0 & = y^2  u'' + y u' + V_y u  \,, \\
\label{eq:Ypot}
 V_y & = \frac{\varpi^2}{4} + \frac{y \zeta (\varpi -\zeta)}{1-y} + \frac{y(\delta^2 + 1/4)}{(1-y)^2} \,.
\end{align}
The form of these equations is identical to the ones derived in the Kerr spacetime~\cite{Yang:2013uba} and the parameters here have the appropriate form in the $Q \to 0$ 
limit\footnote{Although note that Eqs.~(3.11) and~(3.12) of \cite{Yang:2013uba} suffer from two typos: overall, $V_y$ needs to be divided by $2$, and the factor of $y$ is absent from the third term in $V_y$ which involves $\mathcal F_0$, which is the same as the third term here involving $\delta$.}. 

The inner solution is written in terms of hypergeometric functions ${}_2 F_1$,
\begin{align}
\label{eq:NHsln}
u & = y^{-p} (1-y)^{-q} {}_2 F_1 ( \alpha, \beta, \gamma, y)\,,
\end{align}
with 
\begin{align}
p & = i \varpi/2 \,, & q & = -1/2-i\delta \,, \\
\alpha & = 1/2 + i \zeta - i \varpi + i \delta \,, & \beta & = 1/2 -i \zeta + i \delta \,, \\
\gamma & = 1 - i \varpi \,.
\end{align}
Because of the form of this solution and the fact that the outer solution is the same as for Kerr (save for the fact that $\delta$ depends on $a$) the matching proceeds identically to that case. This allows us to calculate the ZDM frequencies for KN.

\subsubsection{Matching and zero-damped modes}
\label{sec:Matching}

The two approximate radial solutions are matched in the region $\sigma \ll x \ll 1$ where both approximations are valid. In this regime the confluent hypergeometric functions simplify, ${}_1 F_1 \to 1$. We must apply an inversion to the hypergeometric function ${}_2 F_1$ using the variable $z = 1-y$, and then take $z \to 0$ as $y\to 1$; this limit essentially takes us to the outer edge of the near-horizon region. 
We equate the two expressions for the radial wave functions, taking care to match the coordinates and the particular wave function used ($u$ versus $R$).
We give further details in Appendix~\ref{sec:MatchingApp}. 
With the outgoing wave condition, the matching can be achieved if the argument of a particular Gamma function is near its poles at the negative integers; specifically we require
\begin{align}
\Gamma[\gamma - \beta] = \Gamma[-n - i \eta] \,,
\end{align}
where $\eta$ is a small correction which guarantees that the matching holds. 
Plugging in all the preceding expressions gives 
\begin{align}
\label{eq:DFfreq}
\omega  = & \, m \Omega_H + \kappa \left[2 m r_+ \Omega_H - \delta - i \left( n + \frac 12\right) + \eta \right] + O(\sigma^2) \notag \\
 =&  \frac{ma}{M^2+a^2} - \frac{M \sigma}{2(M^2+a^2)} \left[\delta +  i \left(n + \frac12\right) \right]  
 \notag \\ &
 + O(\sigma^2, \sigma \eta) \,.
\end{align}
To get to the final line, we used the definition of $\kappa$ and expanded $\Omega_H$ in small $\sigma$.
Our final expression for $\omega$ matches the Kerr result in the limit $Q =0$, where $\epsilon = 1- a/M \ll 1$,
\begin{align}
M \omega_K & = \frac{m}{2} - \sqrt{\frac{\epsilon}{2}} \left[ \delta + i \left(n + \frac 12\right) \right]\,.
\end{align}

For KN black holes, the smaller horizon frequency makes a smaller positive contribution to $\delta^2$ than in Kerr, and it is easier for $\delta^2$ to become negative.
In the Kerr spacetime, the implication of an imaginary value of $\delta$ is the existence of DMs (due to the close relation between $\delta^2 > 0$ and the criterion for a WKB peak outside the horizon, discussed below in~\ref{sec:DampedModes}). An imaginary value of $\delta$ also turns various oscillatory terms in the radial wave function into decaying terms, and suppresses the collective excitation of many ZDM overtones, as discussed in~\cite{Andersson2000,Glampedakis2001,Yang:2013uba}. 
Further consequences include a suppression of multipolar fluxes from a particle orbiting at the innermost stable circular orbit of a nearly extremal Kerr black hole for those multipoles with imaginary $\delta$~\cite{Gralla:2015rpa}, but the full implications of an imaginary $\delta$ have not yet been explored. 
Finally, we note that while usually $\eta$ is extremely small, it can be significant when $\delta$ is very near zero~\cite{Yang:2013uba}.
When $\eta > \sigma$, the $O(\eta \sigma)$ corrections dominate over the $O(\sigma^2)$ terms and the explicit expression for $\eta$ given in Appendix~\ref{sec:MatchingApp} should be incorporated into $\omega$. For even larger values of $\eta$ the matching analysis may break down.
We explore these considerations further in Sec.~\ref{sec:DFNumerics}.

The results presented so far show that ZDMs are present in NEKN in the case of spin-$s$ test fields. We turn next to a discussion of the DMs of these test fields.

\subsection{WKB analysis of the Dudley-Finley equation: Damped modes and spectrum bifurcation}
\label{sec:WKBanalysis}

The DMs cannot be accessed by the nearly extremal matching analysis discussed here, because they violate the assumption that $M k \ll 1$. To see how these DMs behave in the KN spacetime, we require a different tool. 
The WKB analysis of the modes of the DF equation, valid for any $(a,\ Q)$ at high frequencies, provides approximate DF frequencies, gives a criteria for their existence, and provides a unifying picture of the spectrum bifurcation of scalar waves in KN. 
This analysis is a straightforward generalization of the results discussed in~\cite{Yang:2012he}, and was also recently presented in~\cite{Zhao:2015pqa}, which focused on the connection between the WKB frequency formulas and the behavior of unstable null geodesics at the light sphere (see also~\cite{Berti:2005eb}). 
The study~\cite{HodEikonal2012} also extended the WKB results of~\cite{Yang:2012pj,Yang:2013uba} to the case of scalar modes of NEKN, deriving a condition for when the WKB formulas describe ZDMs (a condition also derived in~\cite{Zhao:2015pqa}).

Our focus is on the insights that the WKB analysis provides us in the extremal limit, and we include it here to present a complete picture of the spectrum of scalar modes of NEKN.
We give simplified WKB frequency formulas in Appendix~\ref{sec:WKBApp} not given in~\cite{Zhao:2015pqa}, for reference when we take the nearly extremal limit.
We discuss the WKB formulas for the DMs in this limit, in addition to explicitly deriving the ZDM frequency formula given in~\cite{HodEikonal2012}.
The WKB analysis is brief enough that we include all the essential details here, with some additional formulas in Appendix~\ref{sec:WKBApp}. 

The WKB approximation is an expansion in large frequency. We define the large parameter $L = (l+1/2) \gg 1$, and we also distinguish the real and imaginary parts of $\omega$ as $\omega_R$ and $\omega_I$. With these definitions, the leading order WKB analysis dictates that the eigenvalues of the DF equation have the scalings $A_{lm}\sim O(L^2)$, $\omega_R \sim O(L)$, and $\omega_I \sim O(1)$. These quantities are parametrized in terms of $a,\, Q$, and an ``inclination parameter'' $\mu = m/L$ with $-1 < \mu <1$. These facts, and the dependence of the decay rate on the overtone number $n$ derived in~\cite{Schutz:1985zz}, lead to the definition of convenient rescaled quantities
\begin{align}
A_{lm} = L^2 \alpha(\mu,a,Q)\,, & & \omega &= \omega_R + i \omega_I \,, \notag \\
\omega_R = L \Omega_R(\mu,a,Q)\,, & & \omega_I &= - \left(n + \frac 12\right)\Omega_I(\mu,a,Q) \,.
\end{align}
A key insight drawn from the high-frequency approximation is that the QNMs correspond to rays on the unstable photon orbits of the spacetime~\cite{Goebel1972,FerrariMashhoon1984,Cardoso2009,Dolan2010,Yang:2012he,HodEikonal2012,Zhao:2015pqa}. 
Especially relevant to this viewpoint in the KN spacetime is the work of Mashhoon~\cite{Mashhoon1985}, who studied equatorial null orbits under the assumption that they correspond to the $m=l$ QNMs, although only a full analysis of the wave equation justifies this correspondence~\cite{Yang:2012he,Zhao:2015pqa}. 
As such, the extremum of the radial potential $V_r$ takes on a central role, where it gives the radius of unstable orbits. 
We denote the position of the extremum as $r_0$. 
Although $r_0$ is in fact a minimum of $V_r$, when the problem is recast into form similar to the Schr\"{o}dinger equation, it is $-V_u$ that appears as the potential for the wave function, and so we call $r_0$ the ``peak'' of the potential.

The WKB analysis provides an integral condition for the angular eigenvalues $A_{lm}$, along with algebraic conditions for the real and imaginary part of the frequencies $\omega$. These conditions are the Bohr-Sommerfeld quantization condition
\begin{align}
\int_{\theta_-}^{\theta_+}&d\theta\sqrt{a^2\omega_R^2\cos^2\theta-\frac{m^2}{\sin^2\theta}+A^R_{l m}}=(L-|m|)\pi \,, \\
\sin^2 \theta_\pm &=\frac{2m^2}{A_{l m}+a^2 \omega^{2}_{R}\mp\sqrt{(A_{l m}+a^2 \omega^2_{R})^2+4m^2}} \,,
\end{align}
and
\begin{align}
\label{eq:WKBSolve}
V_u(r_0, \omega_R) = 0\,, & & \left. \partial_r V_u \right|_{r_0,\omega_R} = 0 \,,
\end{align}
where $V_u$ is taken to be the leading order WKB potential
\begin{align}
V_u \approx \frac{K^2 - \Delta \lambda}{(r^2 +a^2)^2} \,.
\end{align}
Generally these conditions must be solved jointly, but the analysis in~\cite{Yang:2012he} reveals that a simple secondary approximation for $A_{lm}$ provides algebraic relations for the mode frequencies which are quite accurate even in the extremal limit. We use this approximation for the WKB analysis of DF, which reads
\begin{align}
\label{eq:AppxA}
\alpha \approx 1 - \frac{a^2 \Omega_R^2}{2}\left( 1- \mu^2 \right) \,.
\end{align}
Then Eqs.~\eqref{eq:WKBSolve} can be reduced to a polynomial equation for $r_0$ and algebraic expressions for $\Omega_R$ and $\Omega_I$ in terms of $r_0$ and the remaining parameters. With $\omega_R$ and $r_0$, the imaginary part of the frequency can be found by evaluating the curvature of the potential at the peak,
\begin{align}
\label{eq:WKBOmegaI}
\Omega_I & = \left. \frac{\sqrt{2 d^2 V_u/dr_*^2 }}{\partial_\omega V_u }\right|_{r_0,\omega_R} \,.
\end{align}
This expression shows that $r_0$ must be a peak of $-V_u$, so that $V_u$ has nonnegative curvature at $r_0$. 

We present the general formulas for $r_0$, $\Omega_R$, and $\Omega_I$ in Appendix~\ref{sec:WKBApp}. We verify in Sec.~\ref{sec:DFNumerics} that these formulas in general have residual errors of order $O(L^{-1})$ and $O(L^{-2})$ respectively, as occurs in Kerr \cite{Yang:2012he}.
Focusing on the extremal and near-extremal cases, we find that the expressions simplify.
As before, we eliminate $Q$ in favor of $\sigma$.

\subsubsection{WKB analysis in the extremal limit}

\begin{figure}[tb]
\includegraphics[width = 1.0 \columnwidth]{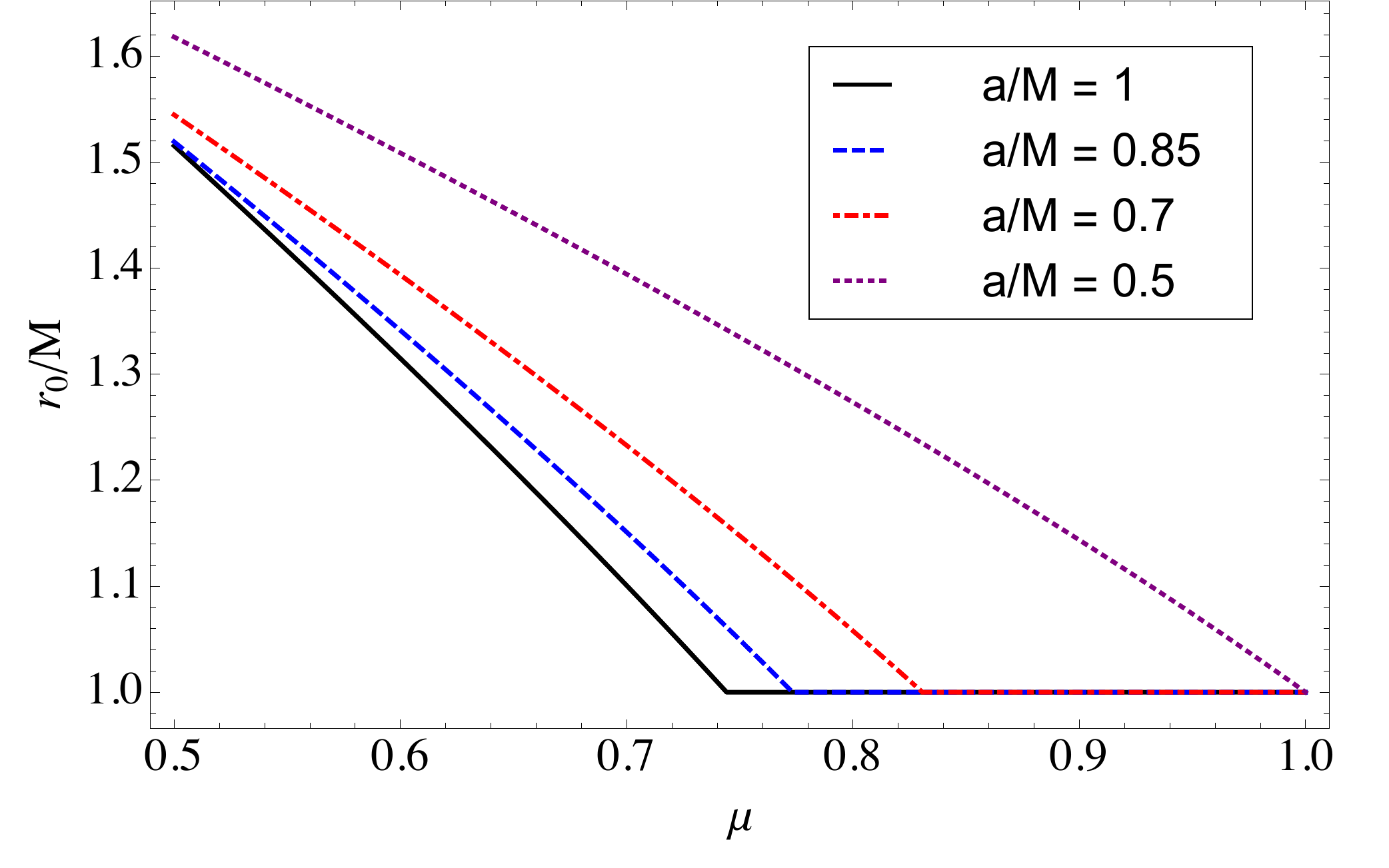} \\
\includegraphics[width = 1.0 \columnwidth]{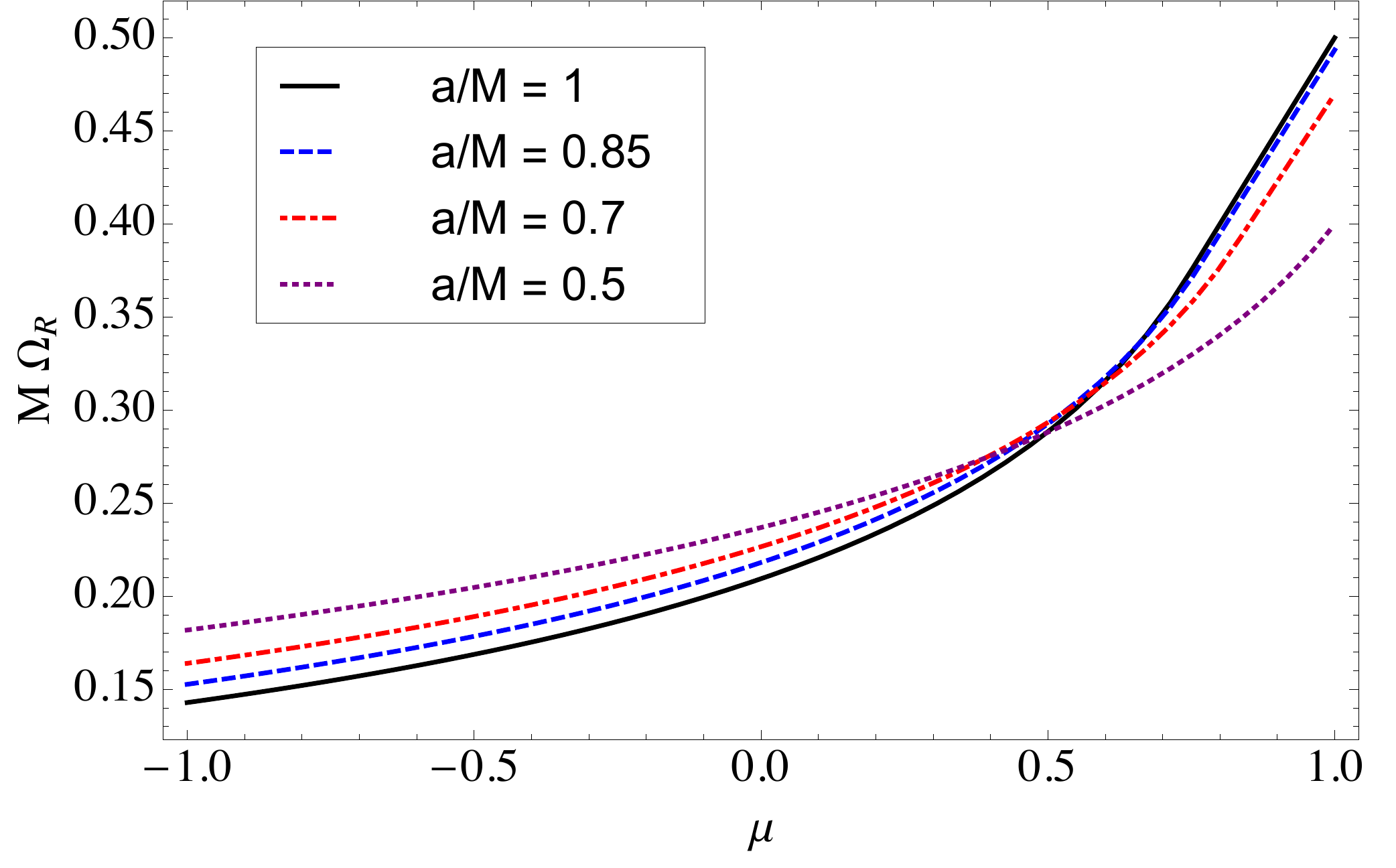} \\
\includegraphics[width = 1.0 \columnwidth]{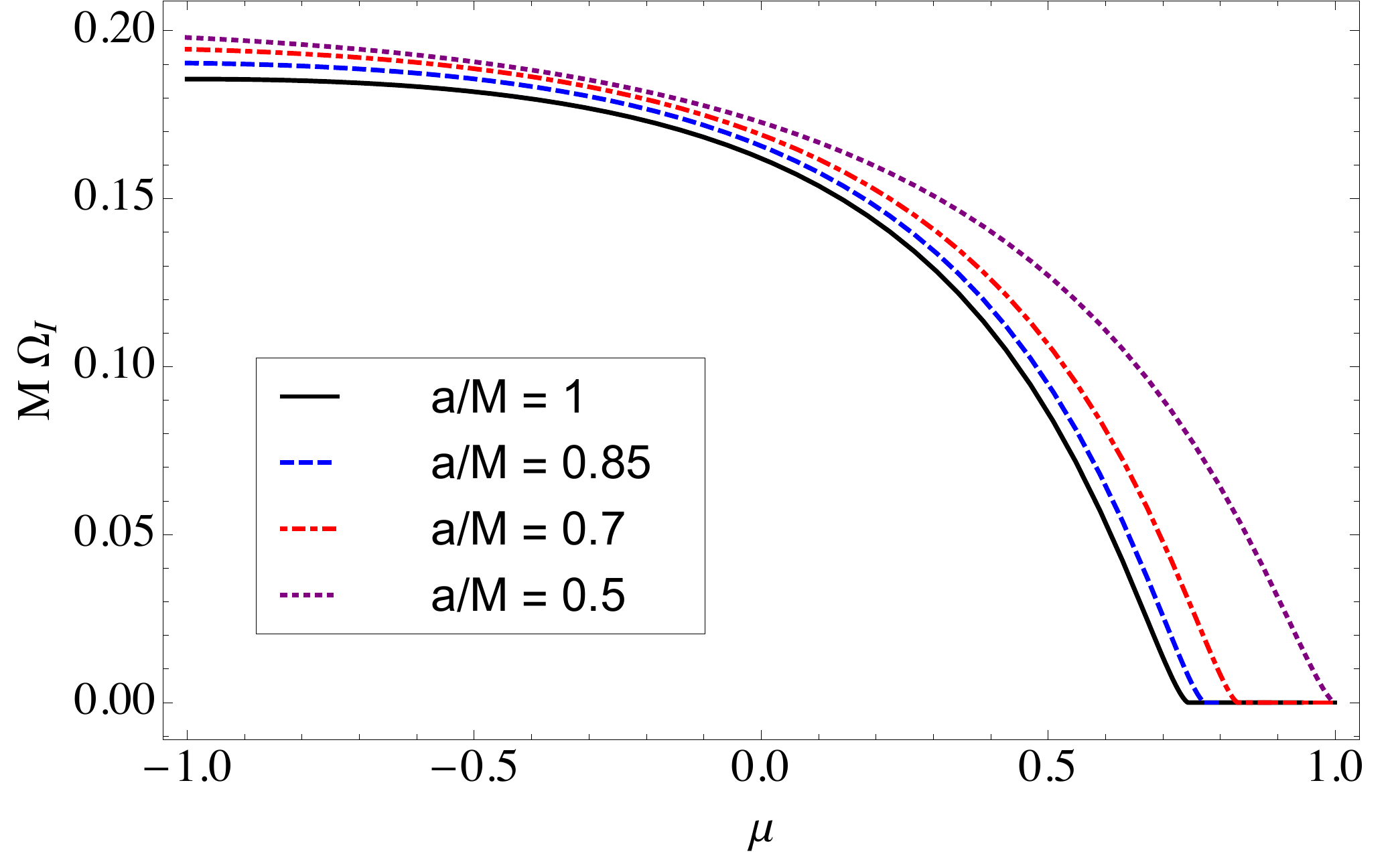}
\caption{WKB quantities plotted against $\mu$ for extremal black holes and various values of $a$. {\it Top panel}: The position of the WKB peak $r_0$. {\it Middle panel}: The WKB frequency $\Omega_R$. {\it Bottom panel}: The WKB decay rate $\Omega_I$.}
\label{fig:ExWKB}
\end{figure}

We focus on the extremal case first, where $\sigma = 0$. Figure~\ref{fig:ExWKB} illustrates $r_0$, $\Omega_R$, and $\Omega_I$ as a function of $\mu$ for a few chosen values of $a$. 
We see that the features found in the WKB analysis of extremal Kerr carry through: at sufficiently high $\mu$, the peak $r_0$ is at the horizon, the frequency asymptotes to $\mu$ times the horizon frequency $\Omega_H$, and the decay rate falls to zero. 
We can understand this behavior by noting that the polynomial for $r_0$ derived from Eqs.~\eqref{eq:WKBSolve} reduces in the limit $\sigma = 0$ to 
\begin{align}
\label{eq:FullExPoly}
(r-M)^2 &[2r^2(r-2M)^2 - 4a^2r(r-2M + \mu^2(r+2M)) \notag \\
&+ a^4 (2 - 3 \mu^2 + \mu^4)] =0\,.
\end{align} 
This has two roots at the horizon, and for sufficiently large $\mu$ there is no other root outside of the horizon. This means that the peak of $V_u$ lies on the horizon, and we can take $r_0 = M$. With this the frequency~\eqref{eq:WKBOmegaR} is
\begin{align}
\Omega_R = \frac{\mu a}{M^2+a^2} \,,
\end{align}
which is the horizon frequency for the extremal black hole.
In addition, $\Omega_I \propto \Delta(r_0)^{1/2}$, so that $\Omega_I = 0 $ when we evaluate it at the horizon radius. 

Meanwhile, when $\mu$ is small enough that an additional root of Eq.~\eqref{eq:FullExPoly} lies outside of the horizon we get a nonzero decay rate even in the extremal limit.
These are the DM frequencies. 
For smaller values of $a$ (larger values of $Q$), larger values of the inclination parameter $\mu$ are required for $r_0$ to occur at the horizon. 
This in turn corresponds to corotating photon orbits which lie closer to the equatorial plane.

Together with our matching results on the existence of ZDMs for all values of $a$ in the nearly extremal case, we see the spectrum bifurcation found in Kerr occurs also for the scalar QNMs of KN. 
An important difference between the KN black hole and Kerr is that even when $\mu = 1$, for sufficiently small values of $a$, DMs exist. 
This is expected: extremal RN black holes, where $a=0$, are known to possess damped scalar modes even for $m=l$.

\begin{figure}[tb]
\includegraphics[width = 1.0 \columnwidth]{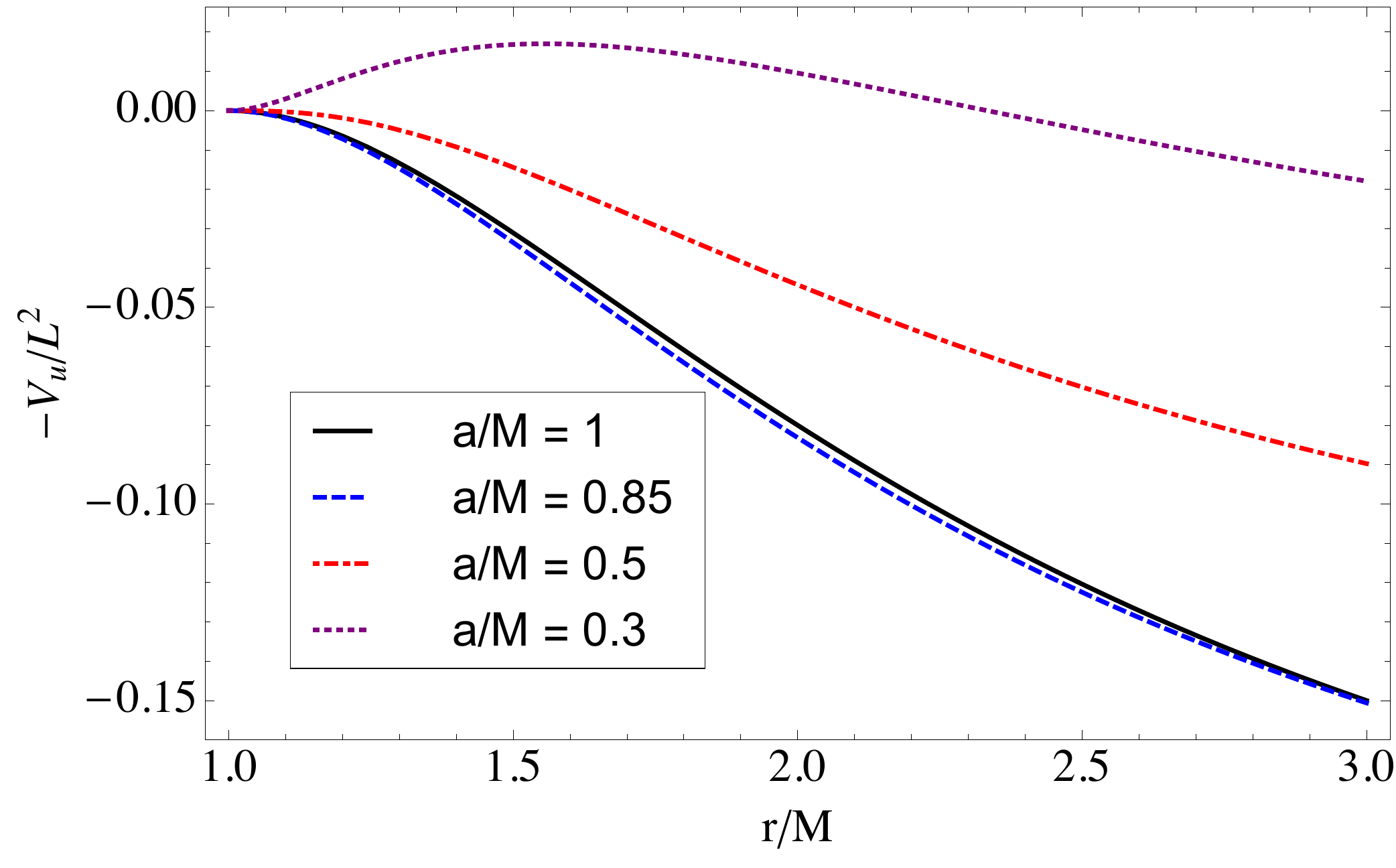}
\includegraphics[width = 1.0 \columnwidth]{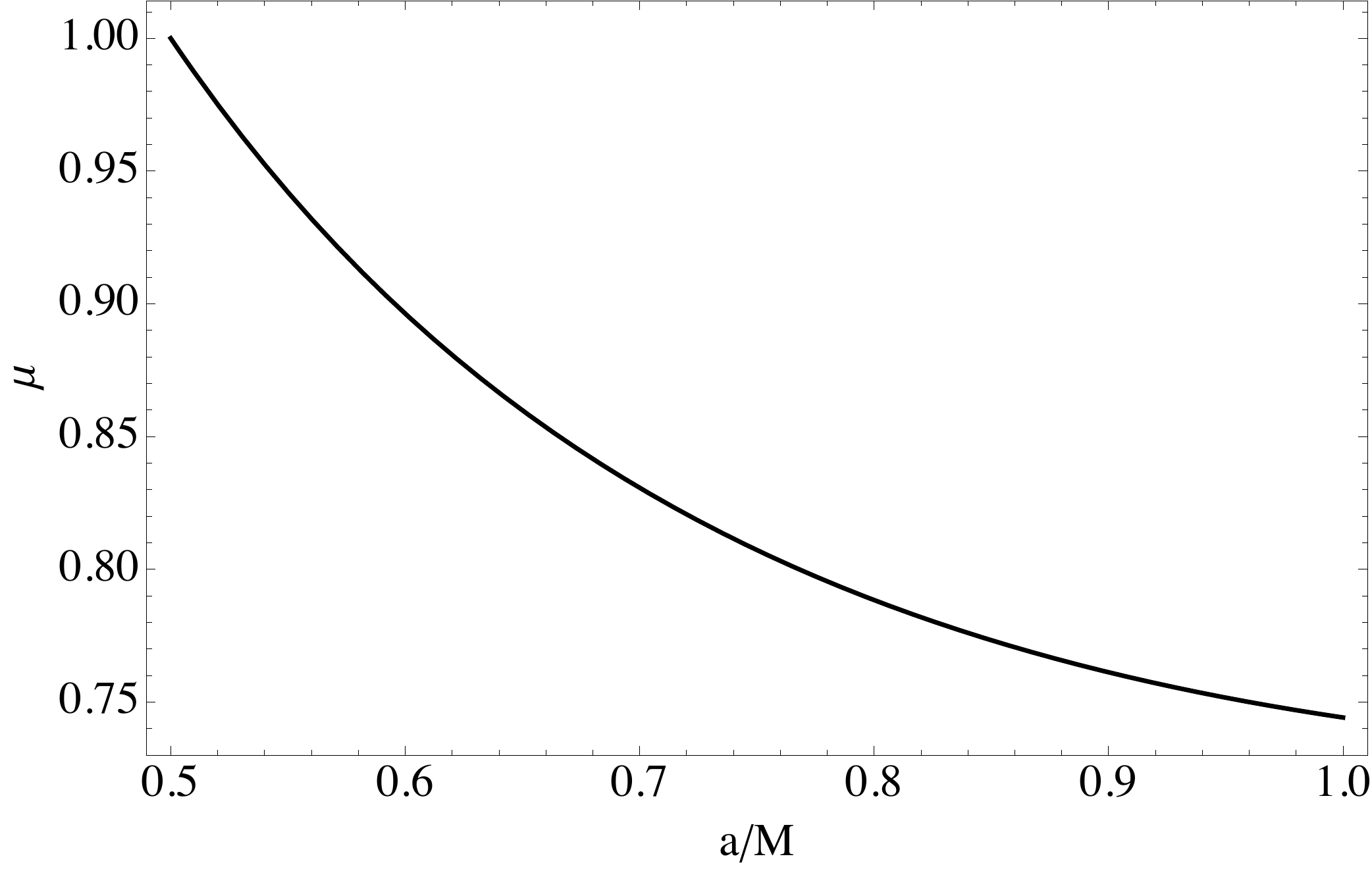}
\caption{{\it Top panel}: Extremal WKB potential $-V_u/L^2$ plotted for various fixed values of $a$. {\it Bottom panel}: The critical inclination parameter $\mu_c$ below which the extremal potential supports a second peak outside the horizon.}
\label{fig:CritMu}
\end{figure}

We can derive an approximate formula for the critical $\mu$ above which the WKB results predict ZDMs by investigating the potential $V_u$ in the extremal case. For this, we set $\Omega_R = \mu \Omega_H$. We know that when $a \to M$ (Kerr), a second peak appears outside the horizon for $\mu < \mu_c \approx 0.74$. For KN, $\mu_c$ is a function of $a$. The top panel of Fig.~\ref{fig:CritMu} shows this extremal potential for $\mu =1$ and various values of $a$. We see the second peak emerge when $a < 0.5$ ($Q > \sqrt{3}/2$). With the above simplifications applied to $V_u$ we find that a peak exists outside the horizon when
\begin{align}
\label{eq:ExPoly}
a^2 \mu^2 (r^2 + 2Mr +a^2 +3M^2) - \alpha(M^2+a^2)^2 = 0
\end{align} 
has a solution for $r>M$. Inserting $r=M$ into the above polynomial, we solve for the critical inclination parameter
\begin{align}
\label{eq:MuCrit}
\mu_c^2 = \frac12 \left(3 + \frac{12 - \sqrt{136 +56(a/M)^2 +(a/M)^4}}{(a/M)^2} \right) \,.
\end{align}
This formula was previously given in~\cite{HodEikonal2012,Zhao:2015pqa}, and it reduces to the known result in Kerr, $\mu_c = [(15 -\sqrt{193})/2]^{1/2}\approx0.74$~\cite{HodEikonal2012,Yang:2013uba}. We plot $\mu_c$ as a function of $a$ in Fig.~\ref{fig:CritMu}. Further setting $\mu_c^2 = 1$ we arrive at the value $a/M = 0.5$ beyond which there are no values of $\mu$ where the WKB peak remains on the horizon.

We see from this that the nearly extremal WKB analysis splits into two cases where we expect simple expressions for the frequencies: when the WKB peak is near the horizon and we have ZDMs, and when the WKB peak is supported away from the horizon we have DMs. We briefly treat each case.

\subsubsection{Zero-damped modes}

Now we consider the case where $\sigma \ll1$, and $\mu >\mu_c(a)$. This is the case where we expect that the WKB approximation describes ZDMs. 
Inspired by the form of Eq.~\eqref{eq:ExPoly} and previous work in Kerr, we define
\begin{align}
\mathcal J^2 & = (m \Omega_H)^2(6M^2 +a^2) - A \,,
\end{align}
so that $\mathcal J^2  > 0$ is the condition for $\mu > \mu_c$ in the extremal limit. 
Next, we make the guess that $r_0$ approaches the horizon at a rate controlled by $\sigma$, $r_0 = M( 1+ c \sigma)$.
The solution for the peak at leading order in $\sigma$ is then
\begin{align}
r_0 \approx M\left(1 + \sigma \frac{M m \Omega_H}{\mathcal J} \right) \,.
\end{align}
For this peak, $\Omega_R$ becomes
\begin{align}
\label{eq:WKBOmegaZDM}
\Omega_R \approx \frac{\mu a}{M^2+a^2} - \sigma \frac{M (\mathcal J/L)}{2(M^2+a^2)} \,.
\end{align}
Finally, inserting these results into the expression for $\Omega_I$ gives
\begin{align}
\Omega_I & \approx \frac{M \sigma}{2(M^2+a^2)}\,.
\end{align}
Collecting these, the WKB approximation for $\omega$ is
\begin{align}
\label{eq:WKBFreq}
\omega & = \frac{m a}{M^2 +a^2} - \frac{M \sigma}{2(M^2+a^2)} \left[ \mathcal J + i \left(n + \frac 12 \right) \right]\,.
\end{align}
Equation~\eqref{eq:WKBFreq} for $\omega$ matches the Kerr limit derived in~\cite{Yang:2012he,Yang:2013uba}. 
In addition, it is the correct WKB limit of the ZDM expression~\eqref{eq:DFfreq} since the only difference between $\delta$ and $\mathcal J$ is at subleading order in $L$.

\subsubsection{Damped modes}
\label{sec:DampedModes}

When the WKB peak is outside the horizon in the extremal limit, the roots of the quartic in the square bracket of Eq.~\eqref{eq:FullExPoly} have involved analytic forms, and the expressions for $\Omega_R$ and $\Omega_I$ do not appear to admit useful simplifications.

The exception is for $\mu = \pm 1$, which gives corotating and counterrotating orbits, respectively. Due to the symmetries of the QNMs, we focus on the case $\mu = 1$ and allow $a$ to vary between positive and negative values, which interpolates between the two cases as $a$ passes through zero. The position of the peak and corresponding frequency are then
\begin{align}
\label{eq:WKBFreqOuterPeak}
r_0 = 2(M-a)\,, & & \Omega_{\rm peak} =  \frac{1}{4M-3a}\,. 
\end{align}
Here we denote the frequency at the WKB peak $\Omega_{\rm peak}$ to distinguish it from the limiting value $\Omega_H$ when both $\mu = 1$ and the peak is at the horizon. These two limits match smoothly when $a = M/2,$ $Q = \sqrt{3} M/2$.
These give a decay rate
\begin{align}
\label{eq:DMExDecay}
\Omega_I = \frac{(M-2a)\sqrt{2a^2-2Ma +M^2}}{\sqrt{2} (M-a)^2(4M-3a)}\,.
\end{align}
This decay rate joins onto the $\Omega_I = 0$ solution as $a \to M/2$.

\subsection{Numerical results for the Dudley-Finley equation}
\label{sec:DFNumerics}

We turn to the problem of determining the accuracy of our analytic results, Eqs.~\eqref{eq:WKBOmegaI} and \eqref{eq:WKBSolve}, and Eq.~\eqref{eq:DFfreq}, valid in the regime of $L\gg1$ and $\sigma\ll1$ respectively. 
We examine the residual errors in these approximations $\Delta\omega \equiv |\omega_\text{A}-\omega_\text{N}|$, where $\omega_\text{A}$ is computed with the appropriate analytic approximation and $\omega_\text{N}$ is computed numerically with sufficiently small error that we can take $\omega_\text{N}$ to be the ``true'' QNM frequency. 
To numerically compute the QNMs we use Leaver's method \cite{Cook:2014cta, Berti:2005eb, Leaver1985, LeaverRN}. 
Leaver's method turns the coupled eigenvalue problem posed by the radial and angular equations \eqref{eq:TeukR} and \eqref{eq:TeukS} into a root finding problem (see~\cite{Cook:2014cta} for a nice discussion).
The eigenvalues, $\omega$ and $A_{lm}$, are reported as simultaneous roots of two infinite, convergent continued fractions, the values of which we denote $\mathcal C^r$ and $\mathcal C^\theta$:
\begin{align}
\label{eq:cfform}
{\mathcal C^r} = \beta_0^r-\frac{\alpha_0^r\gamma_1^r}{\beta_1^r-}\frac{\alpha_1^r\gamma_2^r}{\beta_2^r- \dots} \,.
\end{align}
The indexed greek letters $\alpha^r_i$, $\beta^r_i$, $\gamma^r_i$ are functions of $\omega$, $A$, $a$, $Q$, $l$, and $m$, and the same equation describes $\mathcal C^\theta$ in terms of $\alpha^\theta_i$, $\beta^\theta_i$, $\gamma^\theta_i$. These functions are given in~\cite{Berti:2005eb}.
To implement Leaver's method, $\mathcal C^r$ and $\mathcal C^\theta$ must be truncated, and the resulting expressions are subjected to a numerical root finding algorithm. 
We use $500+$ terms in the continued fractions and \textit{Mathematica's} FindRoot routine. 
For the purposes of analyzing the accuracy of our analytic formula, we ensure that our numerical errors are orders of magnitude smaller than the errors in the analytic approximations.

\subsubsection{Confirmation of the DF WKB results} \label{sec:WKBnum}

\begin{figure*}[tb]
\label{fig:eikonal}
\includegraphics[width =1.0 \columnwidth]{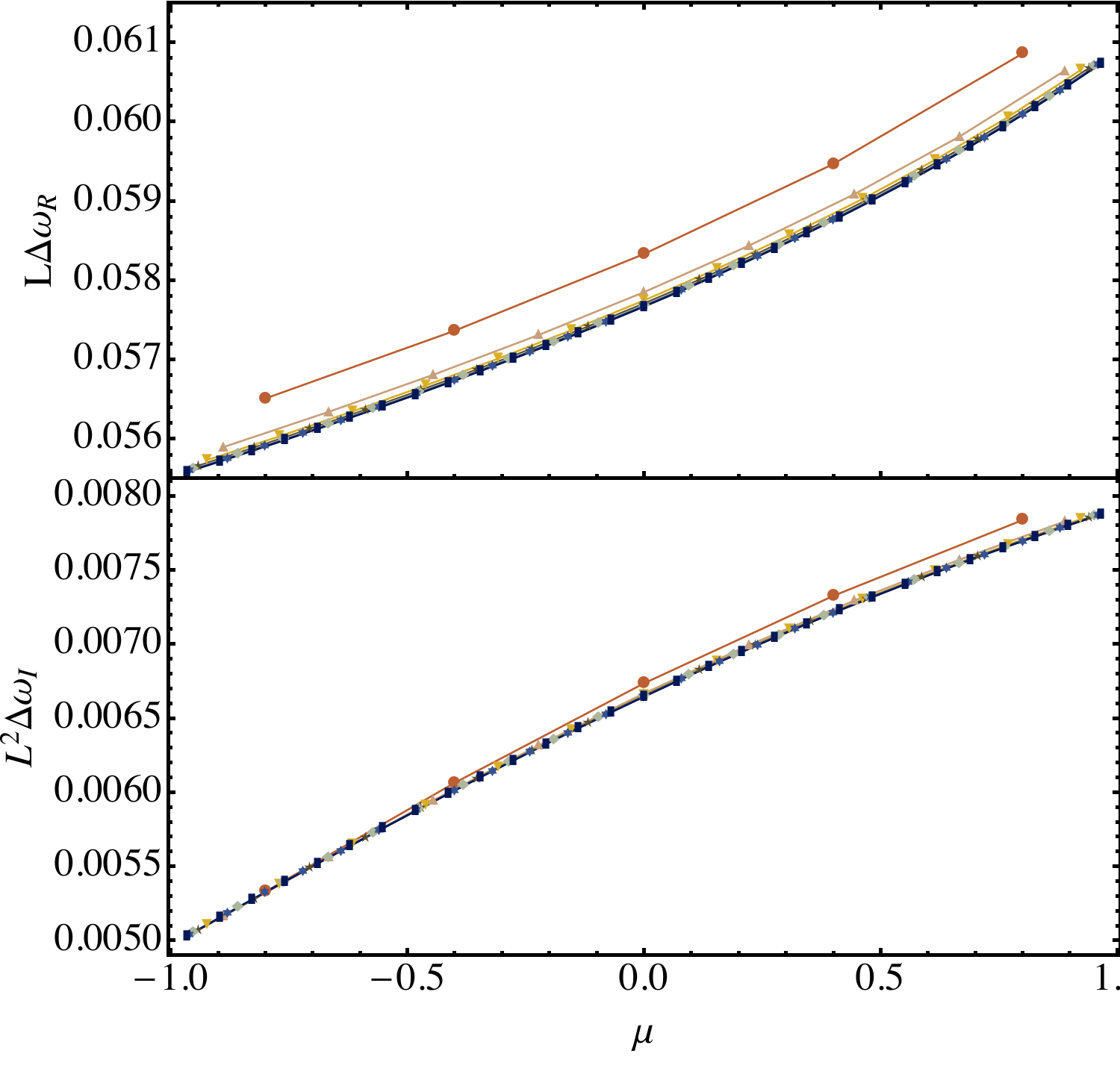}
\includegraphics[width =1.0 \columnwidth]{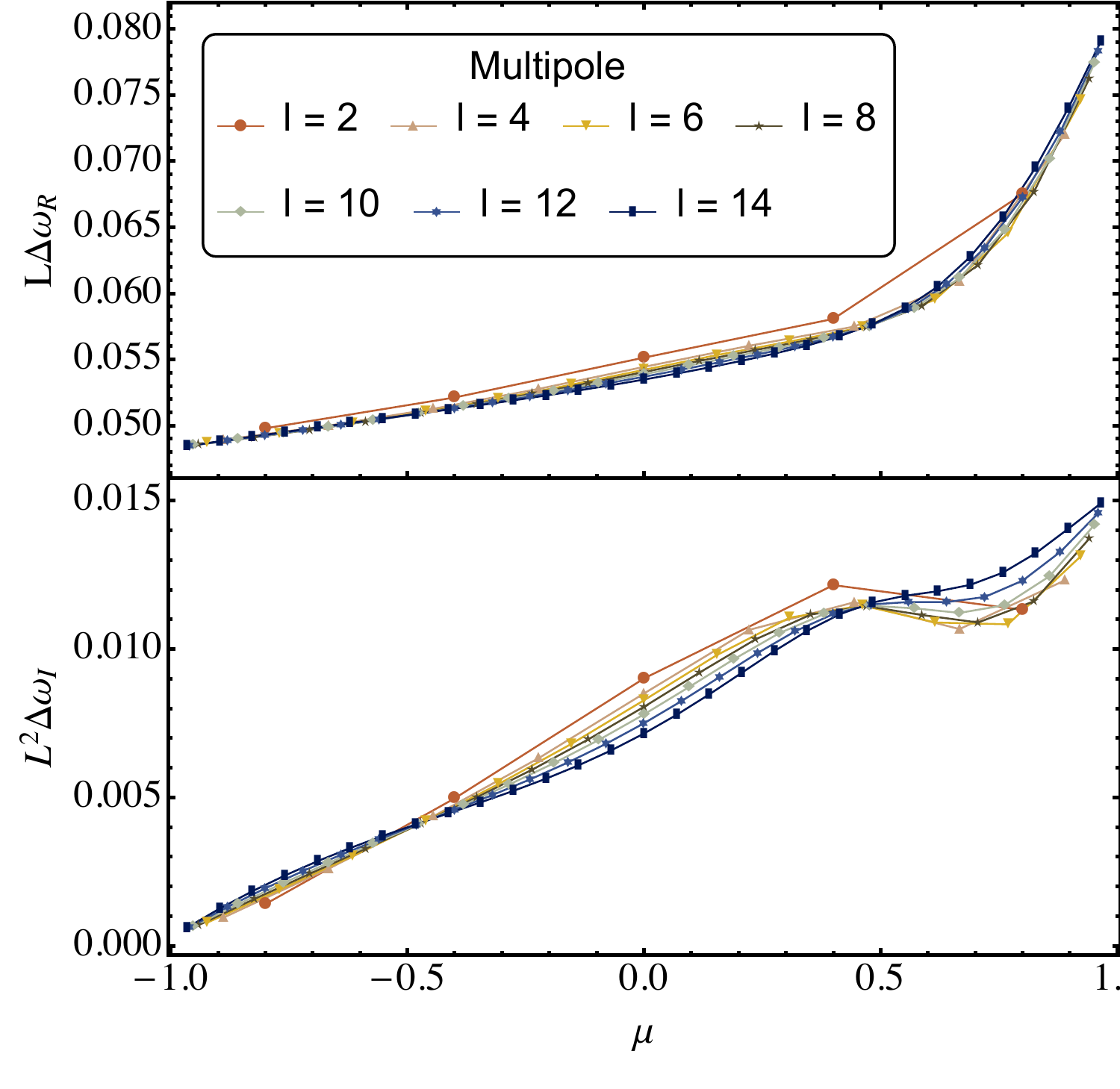} \\
\includegraphics[width =1.0 \columnwidth]{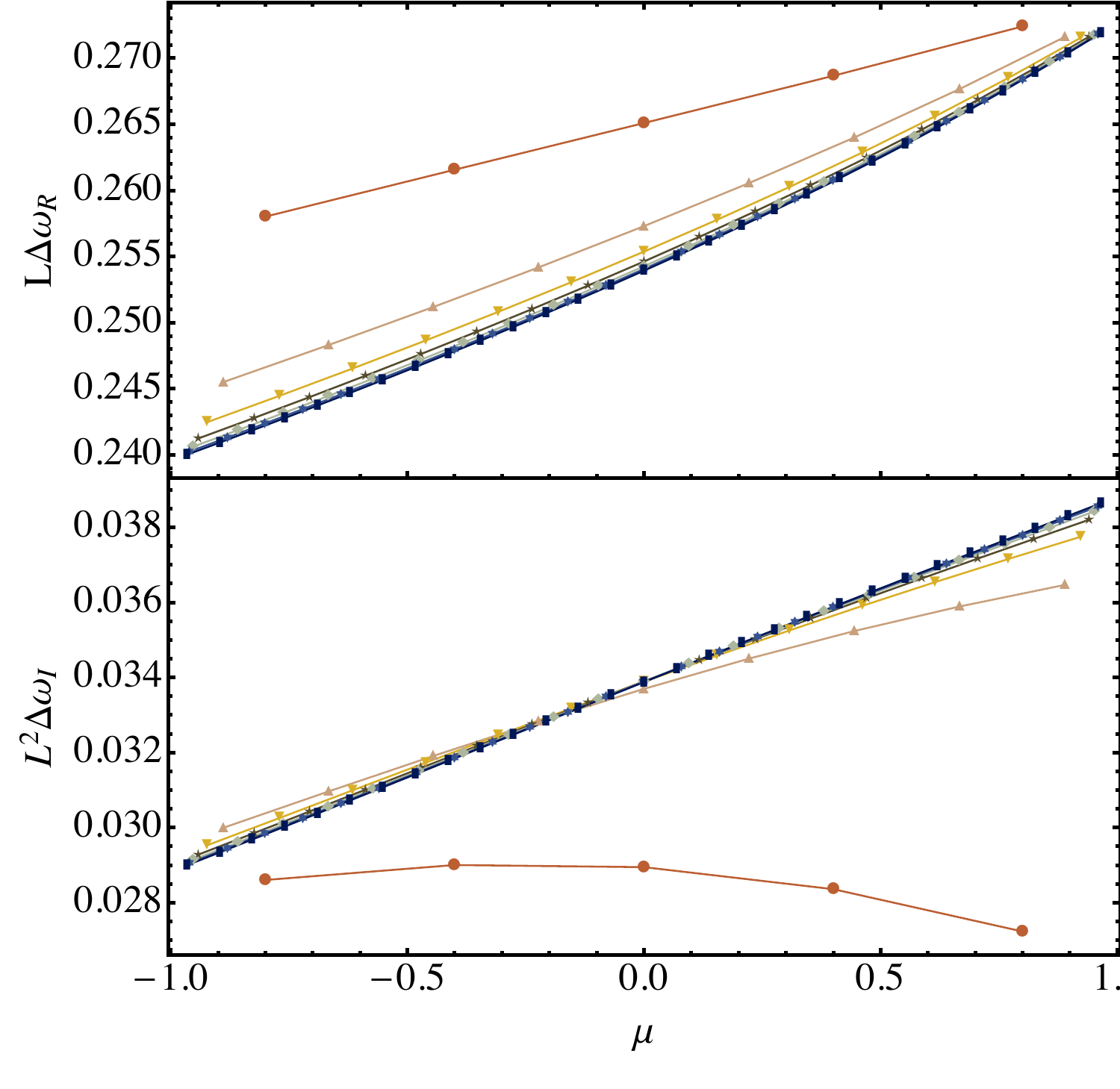} 
\includegraphics[width =1.0 \columnwidth]{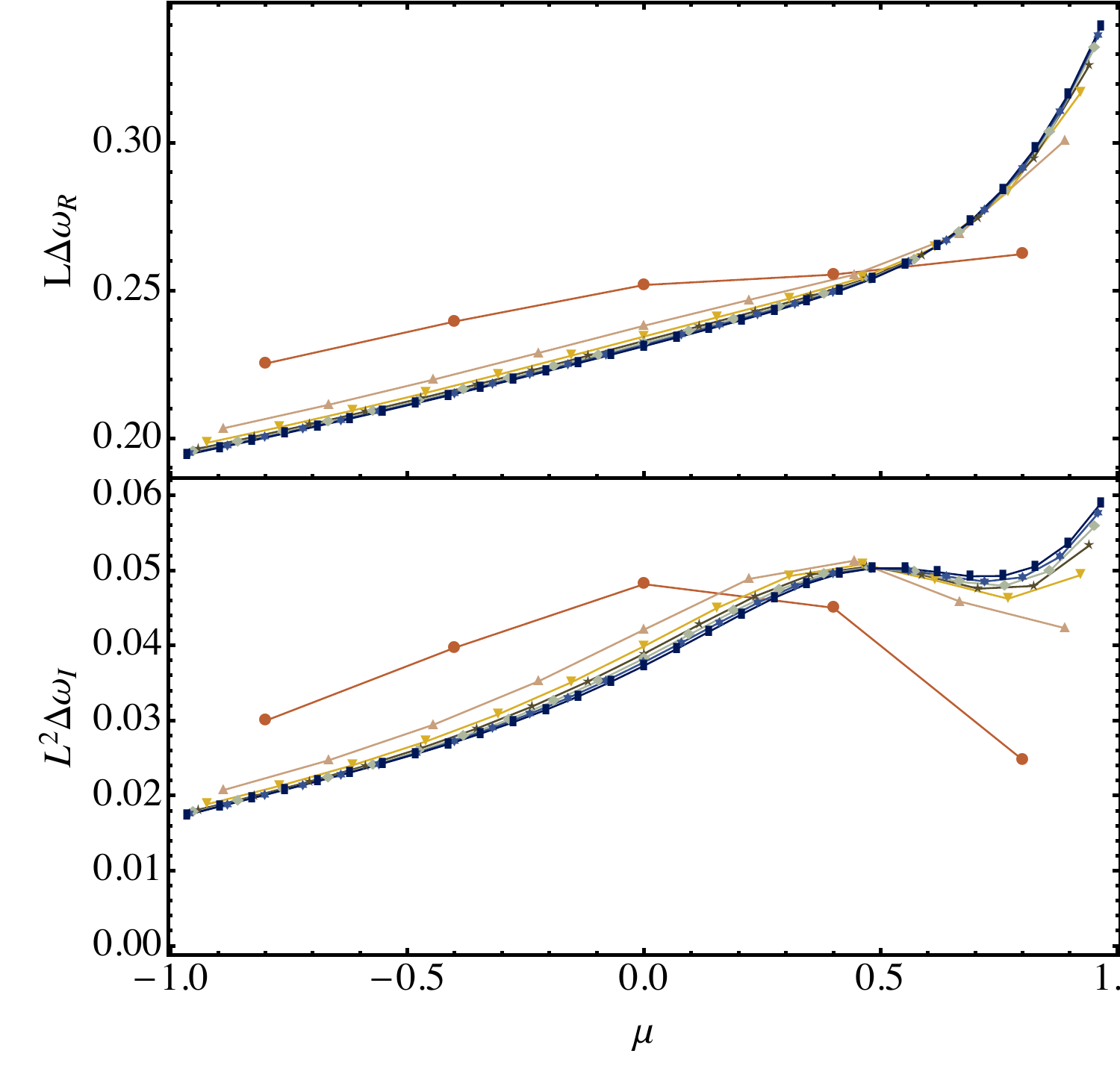}
\vspace{-5mm}
\caption{A numerical study of the error in the DF WKB predictions obtained by solving Eqs.~\eqref{eq:WKBSolve} and~\eqref{eq:WKBOmegaI}. Each panel examines the scaled residual errors $L\Delta \omega_{R}$ and $L^2\Delta \omega_{I}$, for the lowest overtone, using Leaver's method to compute the ``true'' QNM frequency value.
The cases are $Q=0.8M$, $a=0.2M$, $s = 1$ ({\it top left}),  $Q=0.8M$, $a=0.2M$, $s = 2$ ({\it bottom left}), $Q=0.1M$, $a=0.9M$, $s = 1$ ({\it top right}), $Q=0.1M$, $a=0.9M$, $s = 2$ ({\it bottom right}). 
The lines join residuals of constant $l$ and the curves approach a limit curve for every case except  $Q=0.1M$, $a=0.9M$, $s = 1$. For the convergent cases, this indicates the residual error is $O(L^{-1})$ for $\omega_R$ and $O(L^{-2})$ for $\omega_I$. For the case $Q=0.1M$, $a=0.9M$, $s = 1$, the residual errors are still at least $O(1)$ and $O(L^{-1})$, respectively, and are small enough that they may be probing small errors as discussed in Sec~\ref{sec:WKBnum}.}
\label{fig:DFeik}
\end{figure*}

To confirm the WKB predictions for the QNMs of the DF equations, we calculate the analytic $\omega_{\rm A}$ by solving Eqs.~\eqref{eq:WKBSolve} [yielding Eq.~\eqref{eq:WKBOmegaR}] and~\eqref{eq:WKBOmegaI}.
Our closed form expressions assume the approximation for $A_{lm}$ given by Eq.~\eqref{eq:AppxA}. 
When using Leaver's method to find numerical values $\omega_{\rm N}$, we find that seeding the root search at large $l$ is challenging.
To overcome this, we use an accurate approximation for the angular eigenvalue $A_{lm}$ presented in~\cite{Berti:2005gp} (which is especially good at large $l$), leaving only a one-dimensional, numerical root search of $\mathcal C^r$.
In all the cases we have checked, the frequencies calculated in this way are negligibly different than those computed from the coupled root search.

To analyze the error in $\omega_{\rm A}$, we examine modes with $l$ up to $l=14$ for all allowed, discrete values of $\mu=m/(l+1/2)$, with the parameters $Q$, $a$, $s$, and $n=0$ fixed. 
We calculate the scaled residuals $L\Delta \omega_R$ and $L^2\Delta \omega_I$.
These are finite as $L\to\infty$ if $\omega_{\rm A}$ has errors of orders $O(L^{-1})$ and $O(L^{-2})$ in its real and imaginary parts, respectively. 
We join residuals with the same $l$ with lines, so that as $l$ grows the scaled residuals illustrate the limit curve which depends continuously on $\mu$. Four examples are shown in Fig.~\ref{fig:DFeik}, where we increase $l$ from $l = 2$ to $l = 14$.
The parameters for these plots are $s = 1$, $a = 0.2M$, $Q = 0.8M$; $s = 2$, $a = 0.2M$, $Q = 0.8M$; $s = 1$, $a = 0.9M$, $Q = 0.1M$; and $s = 2, a = 0.9M, Q = 0.1M$. We find that the residual errors do generally scale as $\Delta\omega_R = O(L^{-1})$ and $\Delta\omega_I = O(L^{-2})$, which is actually one power of $L$ better than expected by the WKB theory presented in \cite{Schutz:1985zz}. This unexpected accuracy is also seen in Kerr \cite{Yang:2012he}.

It can be difficult to determine visually whether or not the lines are converging to the limit curve.
For the scaling we claim, the spacing between each successive $l$-curve must decrease as they approach the limit curve.
We have checked that this is true for all the cases in Fig.~\ref{fig:DFeik} except for $s = 1$, $a = 0.9M$, $Q = 0.1M$ (top right panels of Fig.~\ref{fig:DFeik}).
There, the spacing between the curves appears to be small and constant.
The magnitude of the residual errors are 10 times smaller than the $s = 2$, $a = 0.9M$, $Q = 0.1M$ case, and we believe it is likely that we are probing errors introduced by using Eq.~\eqref{eq:AppxA} for $A_{lm}$ in the DF WKB implementation, or from using the $A_{lm}$ expansion in the continued fraction $\mathcal C^r$.
Because the residual is quite small, and still an order of $L$ below the leading WKB prediction, we conclude that the WKB formulas here should be accurate enough for most purposes.

\subsubsection{Confirmation of the matched asymptotic expansion results}

\begin{figure*}[t]
\includegraphics[width = 1.0 \textwidth]{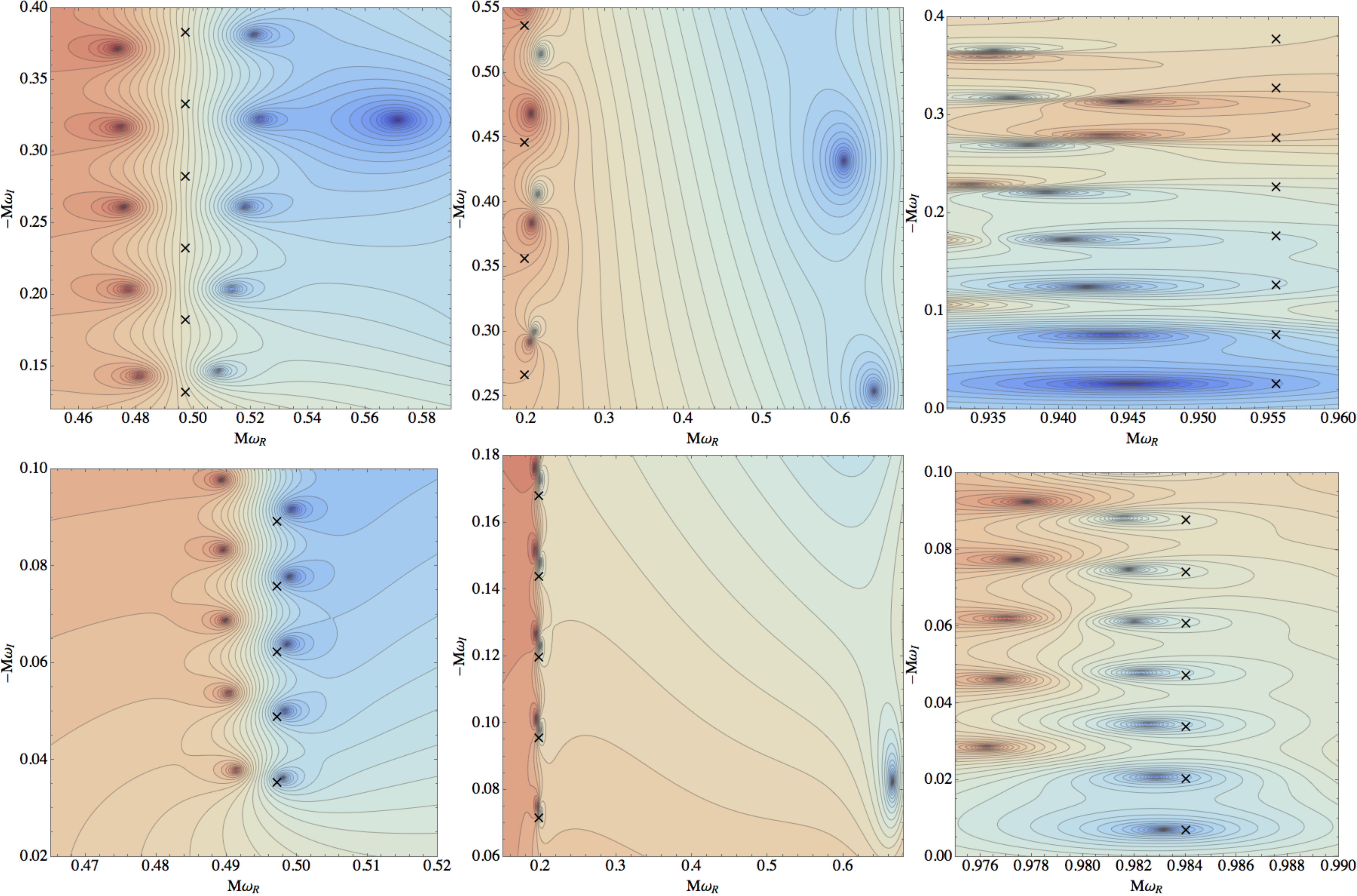}
\caption{Contour plots of the logarithm of Leaver's radial continued fraction $|\mathcal C^r|$.  Blue (dark) areas correspond to smaller values, and redder (lighter) areas to larger values. The zeros of the continued fraction (seen above as a cluster of contours in a dark region) are QNM frequencies, and are usually accompanied by a pole nearby (a cluster of contours in a lighter region).
The predictions from the matched asymptotic expansion are marked by black crosses.
For the top row of panels, we set $\sigma = 0.182$ and for the bottom row we set $\sigma  = 0.049$. 
{\it Left column}: The case $a = 0.9 M$, $s = 0$, $l=2$, and $m =1$. The zeros appearing in a vertical line with $M\omega_R \approx 0.5 $ correspond to to ZDMs. A damped mode is also visible to the right of the ZDMs in the top panel. In the bottom panel, the ZDMs stack more neatly and have moved closer to the real axis; the DM is outside of the range of the plot.
{\it Center column}: The case $a = 0.1 M$, $s = 0$, $l=2$, and $m =2$. Again, some of the DMs are visible on the right.
{\it Left column}: The case $a = 0.9 M$, $s = 0$, $l=2$, and $m =2$. As $\delta^2>0$, there are no damped modes in the spectrum.}
\label{fig:cfplot}
\end{figure*}

In this section we investigate the scalar ZDMs $(s=0)$ of the KN spacetime and compute $\omega_{\rm A}$ using Eq.~\eqref{eq:DFfreq}, obtained from the matched asymptotic 
expansion\footnote{We note that charged, massive scalar QNMs were investigated using Leaver's method in~\cite{KonoplyaNEKN}. That study provided some results in the massless, uncharged limit, which is the problem we investigate here. However, that study found that in the extremal limit, QNMs which were ZDMs in Kerr limited to a finite decay rate for $Q \neq 0$. This is in conflict with the results we present here.}. 
We verify that the residual error in the analytic formula scales as $\sigma^{2}$, which can be taken as an independent check of the validity of the matched asymptotic calculation. 

In certain regions of parameter space, it can be difficult to apply Leaver's technique because $\mathcal C^r$ becomes a rapidly varying function of $\omega$ and the success of the root-finding scheme becomes heavily dependent on the accuracy of the initial seed. 
We find in practice that this can occur when many QNM frequencies bunch together, which is the qualitative behavior we expect for the ZDMs. To get a sense of where the roots are, we borrow a technique from~\cite{Yang:2012pj} where contours of the logarithm of $|\mathcal C^r|$ are plotted as a function of complex $\omega$. 
QNM frequencies appear as clusters of contours forming circles around places where $\mathcal C^r$ is zero. 
Nonphysical poles~\cite{LeaverPoles} of $\mathcal C^r$ also appear as clusters of contours forming circles; however these can be distinguished by examining the value of the continued fraction. 
In our plots, blue (dark) regions correspond to smaller values of $|\mathcal C^r|$, while red (light) regions correspond to  larger values. 

Such plots can immediately demonstrate the existence of separated families of DMs and ZDMs. In these plots we fix $a$, $s=0$, $l$, and $m$, and examine two different values of $\sigma$. The ZDMs appear in a roughly vertical line with $\omega_R \approx m \Omega_H$. In the more extreme case, the line shifts toward the real axis and stacks more neatly, as seen in the left column of Fig.~\ref{fig:cfplot} where $l=2$, $m=1$, and $a = 0.9 M$. DMs can be distinguished in these figures, and move only slightly as $\sigma$ is decreased. A single DM can be seen in the top left panel of Fig.~\ref{fig:cfplot}.

\begin{figure}[tb]
\includegraphics[width =.95 \columnwidth]{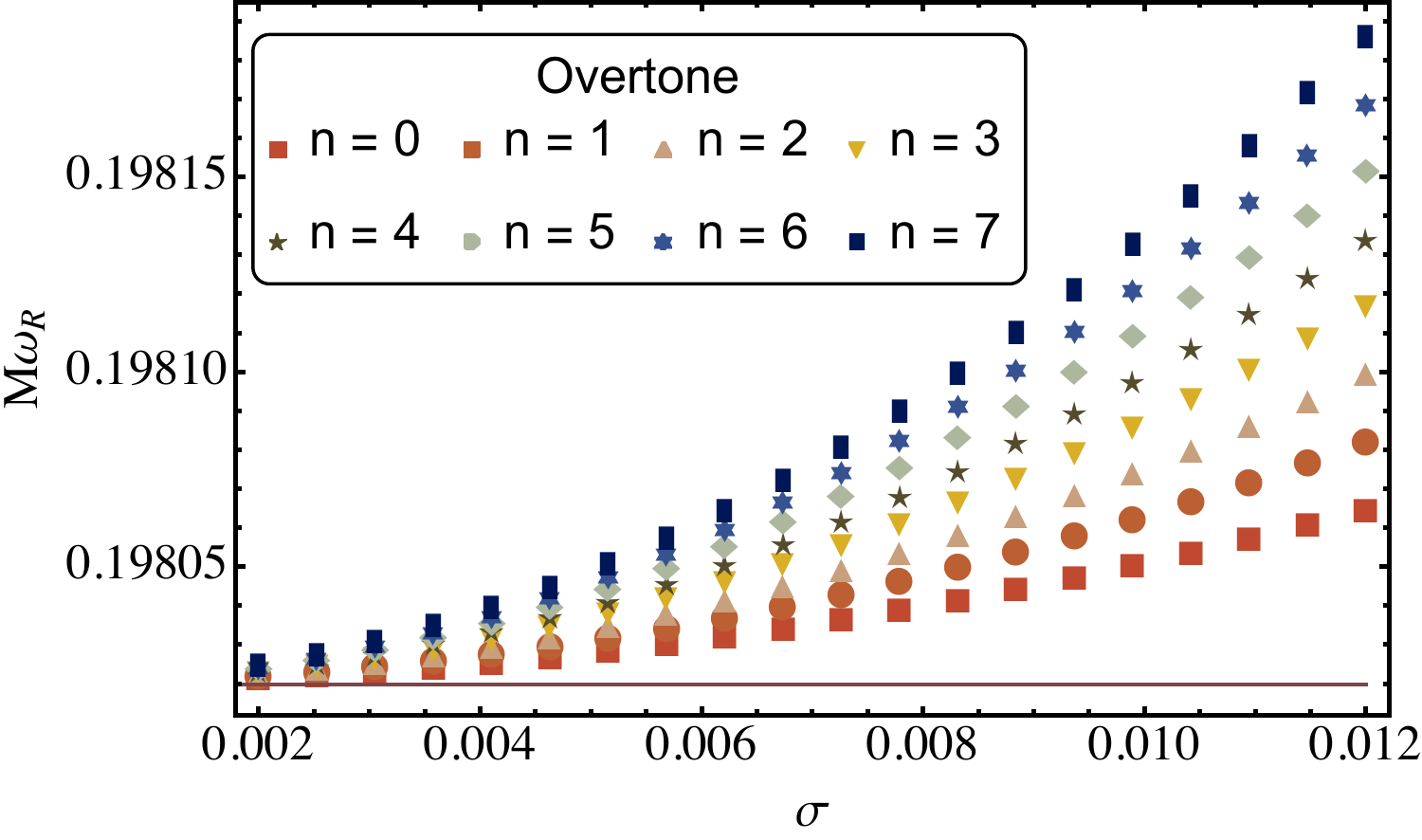}
\includegraphics[width =.95 \columnwidth]{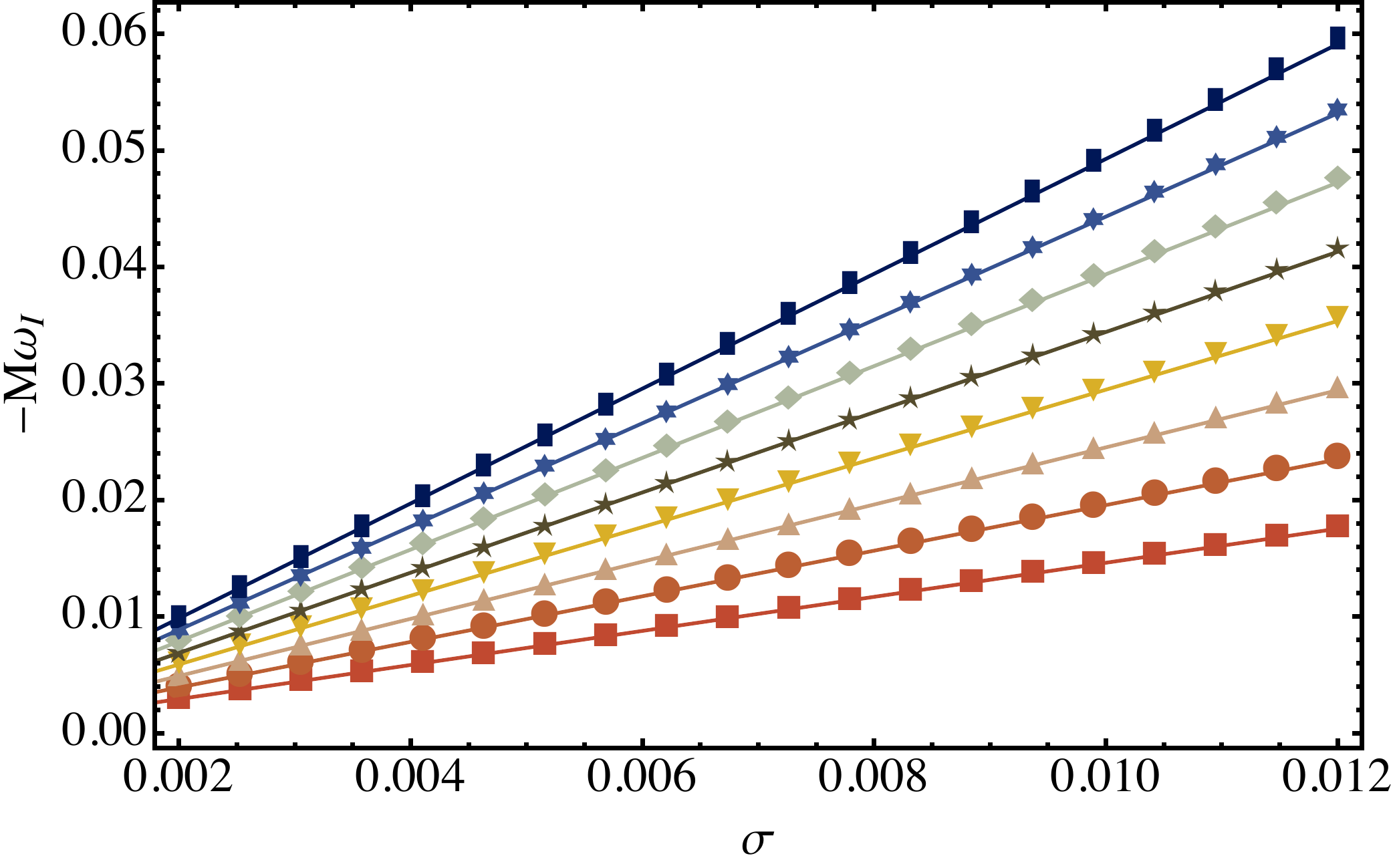}
\caption{{\it Top panel}: We fix $a = 0.1 M$, $s=0 $, $l=2$, $m =2$ and study $\omega_R(\sigma)$ when $\delta^2 <0 $. Numerical calculations using Leaver's method appear as points and the analytical prediction is the solid line. The analytical prediction for $\omega_R$ is independent of $n$ and $\sigma$ at order $O(\sigma)$, so that the errors are $O(\sigma^2)$. {\it Bottom panel}: Fixing the same parameters, we study $\omega_I(\sigma)$. 
}
\label{fig:a1res}
\end{figure}

\begin{figure}[tb]
\includegraphics[width =.95 \columnwidth]{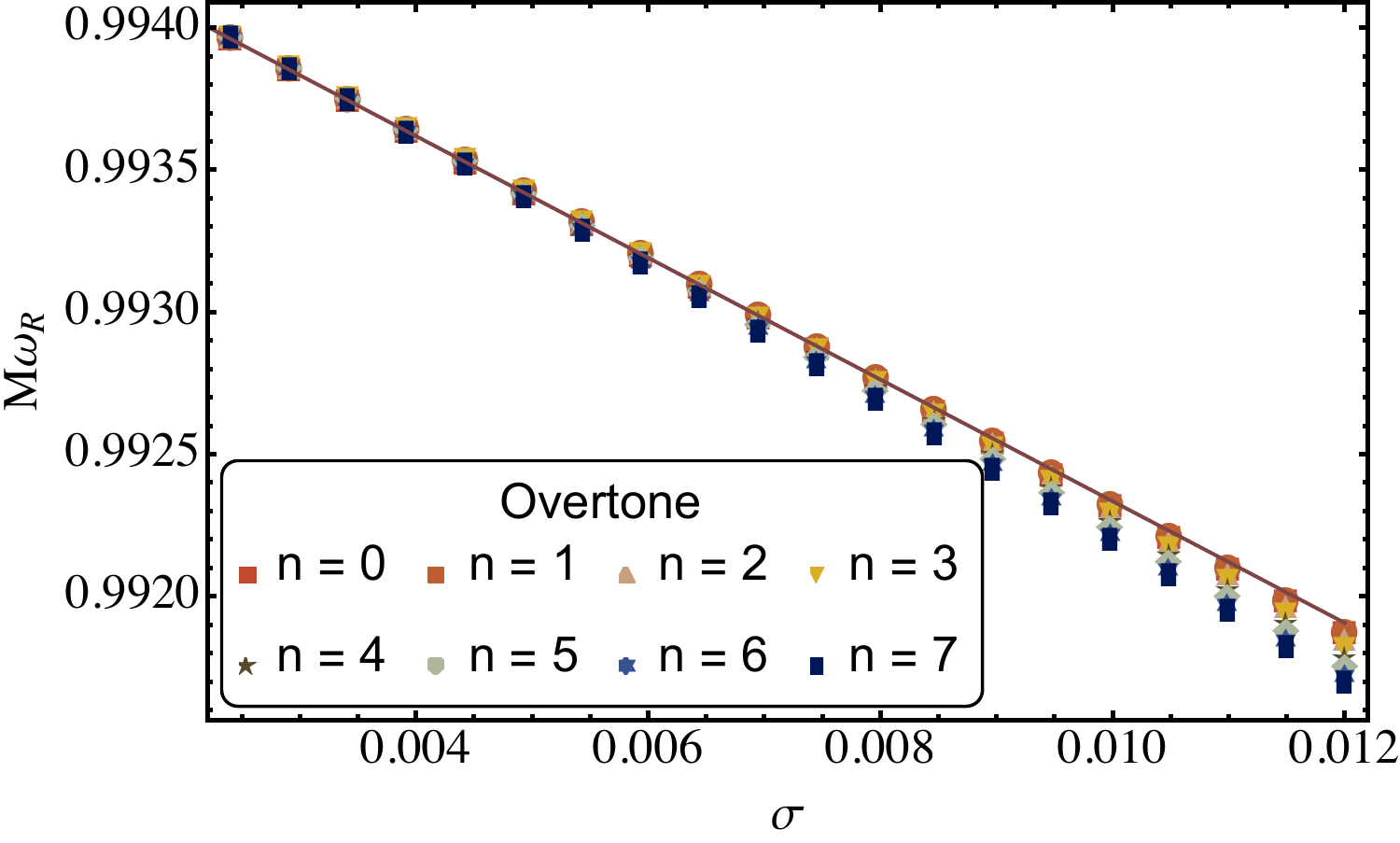}
\includegraphics[width =.95 \columnwidth]{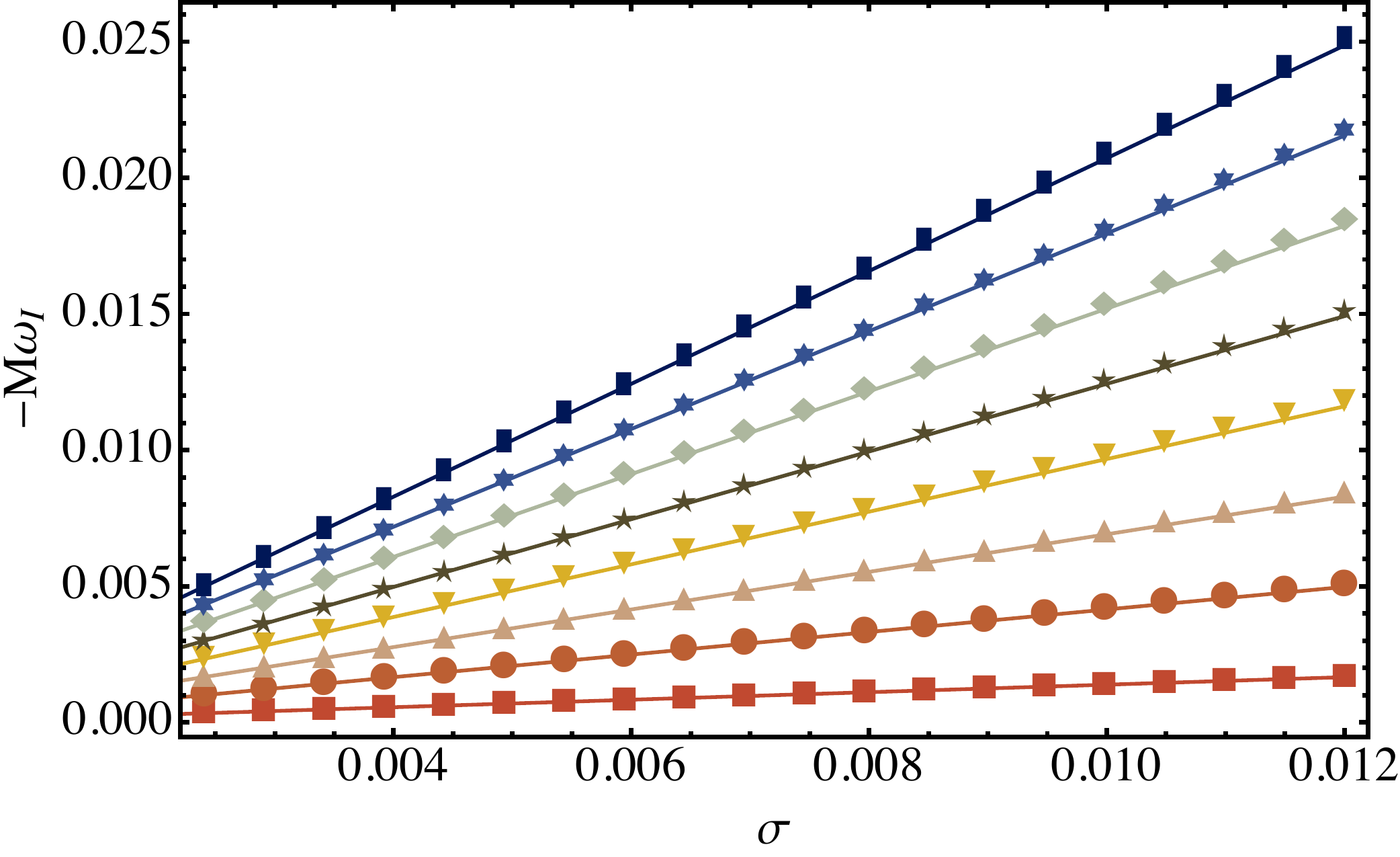}
\caption{{\it Top panel}: We fix $a = 0.9 M$, $s=0 $, $l=2$, $m =2$ and study $\omega_R(\sigma)$ when $\delta^2 > 0$. Numerical calculations using Leaver's method appear as points and the analytical prediction is the solid line. Unlike the $\delta^2<0$ case, the analytic prediction for $\omega_R$ now has linear $\sigma$ dependence. {\it Bottom panel}: Fixing the same parameters, we study $\omega_I(\sigma)$.}
\label{fig:a9res}
\end{figure}

To illustrate the accuracy of our analytic approximations, we fix $s = 0$, $l = 2$ and test Eq.~\eqref{eq:DFfreq} for chosen values of $m$ and $a$. Together with a choice of $\sigma \ll1  $, this fixes $Q$.
Since the QNMs with $\delta^2>0$ are expected to be qualitatively different from those with $\delta^2<0$, we choose a value of $a$ covering each case.
We start with $\delta^2<0$, and examine QNMs with $a = 0.1 M$, $m =2$  while varying $\sigma$. The center column of Fig.~\ref{fig:cfplot} presents the contour plots for two values of $\sigma$, and we observe the qualitative signatures of ZDMs. 
Figure~\ref{fig:a1res} contains a more detailed look at the $\sigma$-dependence of $\omega$ for the eight lowest overtones, and demonstrates the excellent agreement with the analytic formula. 
For the $\delta^2>0$ regime, we fix $a = 0.9 M$ and $m =2$. In Figs.~\ref{fig:cfplot} (right column) and \ref{fig:a9res}, we present the similar plots to Figs.~\ref{fig:cfplot} (center column) and \ref{fig:a1res}, except with $a=0.9M$. Again we observe ZDMs in good agreement with the analytic prediction.

\begin{figure*}[tb]
\includegraphics[width =1.0 \columnwidth]{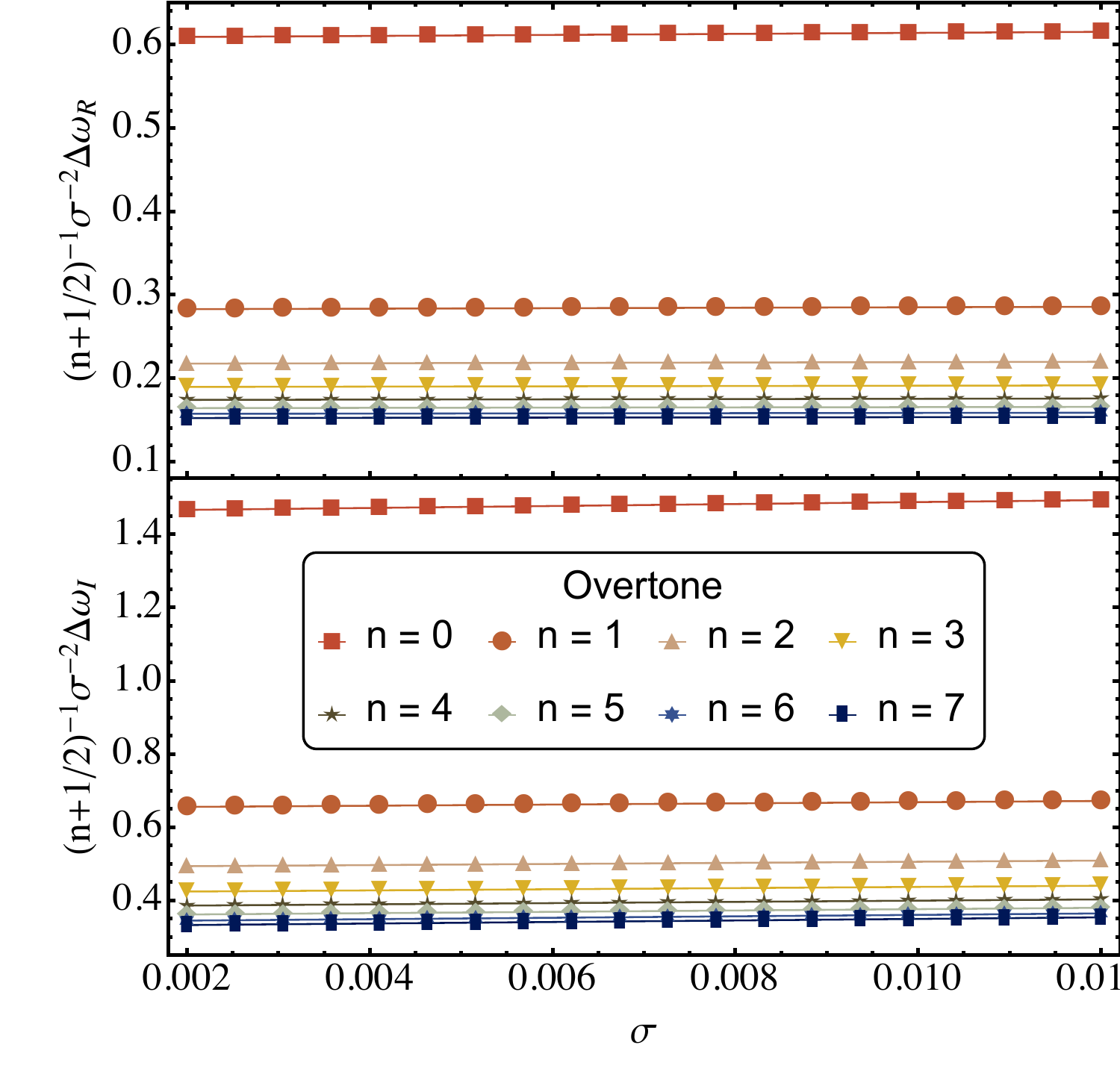}
\includegraphics[width =1.0 \columnwidth]{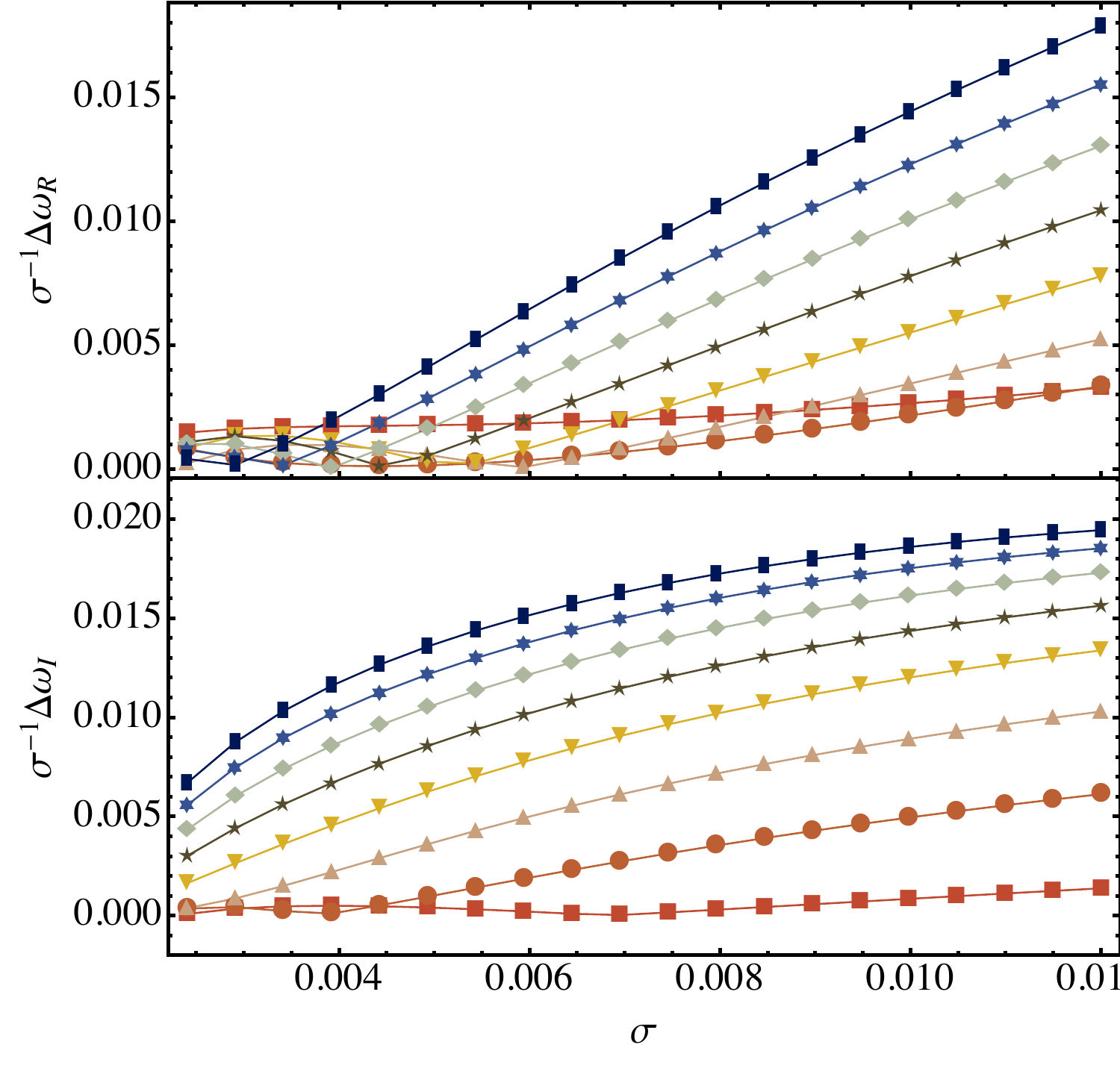} \\
\vspace{-5mm}
\caption{Scaled residual errors of the $s=0$ ZDM frequencies for the first eight overtones of a KN black hole. 
{\it Left}: We study the $\delta^2<0$ regime and fix $a  = 0.1 M$, $l=2$, $m =2$, letting $\sigma$ vary. We find that the residuals are $O(\sigma^2)$. {\it Right}: We study the $\delta^2>0$ regime and fix $a  = 0.9 M$, $l=2$, $m =2$, letting $\sigma$ vary. The $O(\sigma)^2$ scaling of the residuals ceases to hold at low enough $\sigma$, as discussed in the text.}
\label{fig:DFsigcon}
\end{figure*}

We expect the analytic formula to have residual errors of $O(\sigma^2)$. 
Hence we expect the quantity $\sigma^{-2}\Delta \omega$ is $O(1)$ as $\sigma \to 0$, representing the next order in the nearly extremal expansion. 
In the left panel of Fig.~\ref{fig:DFsigcon}, we return to the $\delta^2 <0$ case $a = 0.1M$, $l = m =2$. We plot the scaled residuals errors $M (n+1/2)^{-1}\sigma^{-2}\Delta \omega$ for the real and imaginary parts of $\omega$ of the lowest eight overtones as we vary $\sigma$.
For each overtone, we can follow the curve from right to left and observe that the scaled residuals become constant, demonstrating $M\Delta\omega = O(\sigma^{-2})$. 
We can also follow the curves from top to bottom and observe that they cluster around a limit curve, indicating $M\Delta \omega=O(n+1/2)$ at large $n$. 

In the right panel of Fig.~\ref{fig:DFsigcon}, we return to the $\delta^2 >0$ case $a=0.9 M$, and $l = m =2$. 
We observe that the residual errors scale are $O(\sigma)$, since the quantity $\sigma^{-1}\Delta \omega$ approaches a nonzero finite number as $\sigma \to 0$.  
In our case studies, all of the modes with $\delta^2 >0$ had residual errors one power larger than the modes with $\delta^2 < 0$ at these small values of $\sigma$.
This indicates that in these cases, where only ZDMs are present, the additional term $\eta$ in Eq.~\eqref{eq:DFfreq} (discussed further in Appendix \ref{sec:MatchingApp}) is not completely negligible, with $\eta \sim 10^{-3}$.
When $\sigma \sim 10^{-3}$, the $O(\eta\sigma)$ correction is not negligible relative to the $O(\sigma^2)$ term in  Eq.~\eqref{eq:DFfreq} and the $O(\sigma^2)$ convergence is not seen.

\begin{figure}[tb]
\includegraphics[width =1.0 \columnwidth]{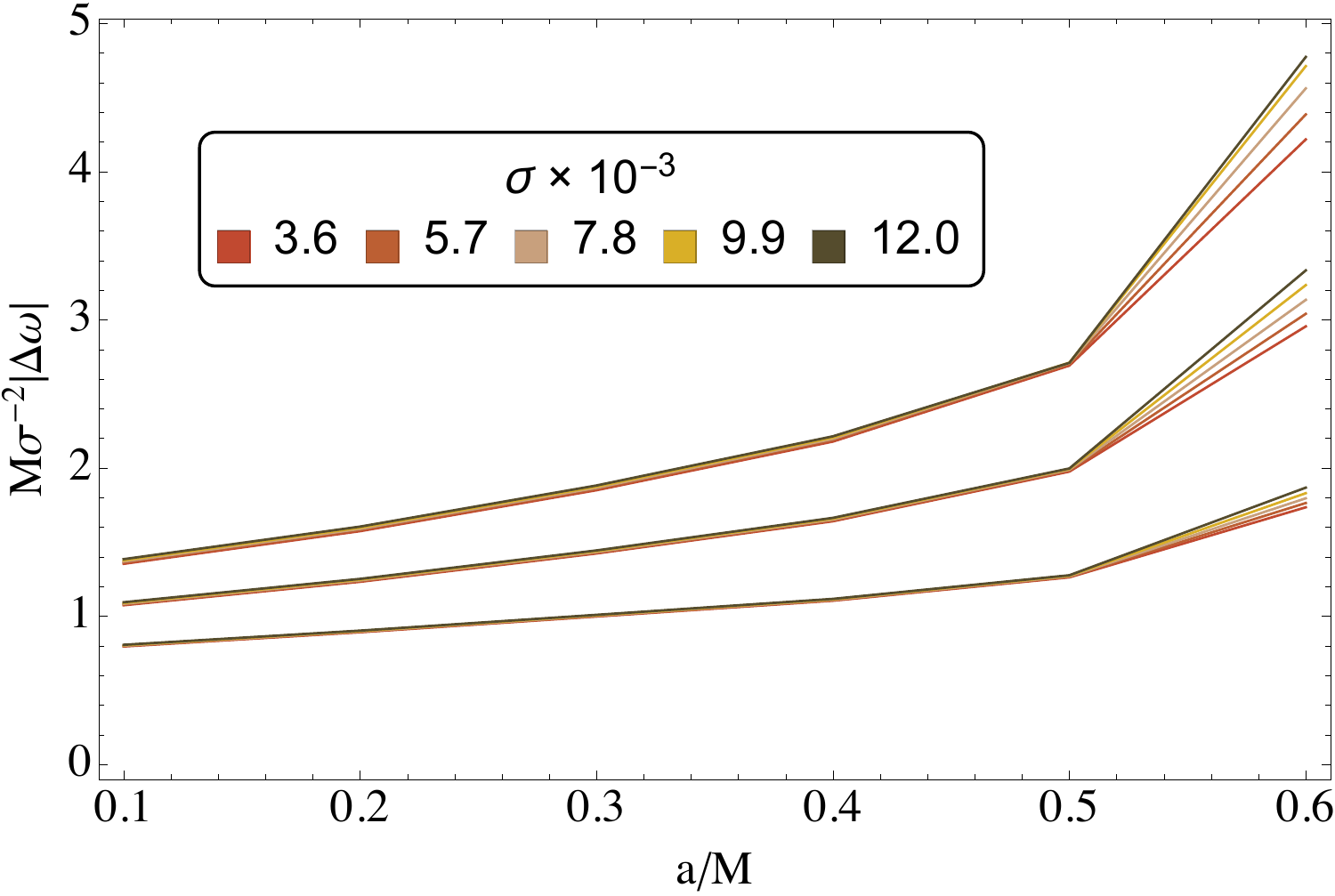}
\caption{An evaluation of the convergence of the ZDM frequency formula Eq.~\eqref{eq:DFfreq}. We plot $\sigma^{-2}|\Delta\omega|$ for the three lowest overtones for each of six values of $a$, with $s=0$, $l=2$, $m=2$. In each case, $\delta^2<0$, and $\sigma$ is decreased towards extremality. The overtones are distinguished by the fact that $|\Delta\omega|$ increases with overtone. A finite value of the limit corresponds to following the curves in each overtone band from top to bottom and observing a linear approach to a limit curve. We have checked that the points linearly approach a limit curve for each value of $a$ presented here.}
\label{fig:errscale}
\end{figure}

Meanwhile, in Fig.~\ref{fig:errscale} we show that $\eta$ is so small that $\Delta\omega = O(\sigma^{-2})$ in practice when there are DMs ($\delta^2<0$). 
Here we first fix a value of $a$ and $n$ and calculate  $M (n+1/2)^{-1}\sigma^{-2}\Delta \omega$ for several values of $\sigma$, spaced by roughly $\Delta \sigma \approx 2 \times 10^{-3}$. 
We plot these points in Fig.~\ref{fig:errscale} above the corresponding value of $a$. 
For each of the three overtones, the lines corresponding to the same value of $\sigma$ are plotted with the same color and the limit  $\mathcal \sigma \to 0$ is taken by following the curves from top to bottom. 
We plot the data for six values of $a$, evenly spaced from $a = 0.1 M$ to $a = 0.6 M$, and connect  data points with the same values of $\sigma$ to allow for a rough interpolation to other values of $a$.
The exception is $a = 0.7M$ (not shown), where $\delta^2$ is negative and close to zero, and we expect a larger value of $\eta$. In this case, we find that the residuals do not scale as $\sigma^{2}$.

Overall, the numerical results indicate that the simple expression~\eqref{eq:DFfreq} can be used over a large range of the parameter space. However, care must be taken to include the correction $\eta$ when $\delta^2$ is close to zero, and when the hole is very close to extremality, e.g. $\sigma \sim 10^{-3}$ (when $Q = 0$, $\sigma \sim 10^{-3}$ gives $a/M \sim 1 - 10^{-6}$).

\section{Gravito-electromagnetic modes of Reissner-Nordstr\"{o}m}
\label{sec:RN}

The existence of ZDMs for spin-weighted scalar perturbations of NEKN black holes naturally raises the question of whether ZDMs also exist for gravitational and electromagnetic perturbations of KN.
As discussed previously, the equations for GEM perturbations of KN are coupled, and so electromagnetic perturbations cannot be considered separately from gravitational perturbations.
We can begin to approach this challenging problem by considering first the simpler case of RN.
The results of Sec.~\ref{sec:DF} hold for any spin parameter $a$, including the limit of $a \to 0$, and this shows that even in the well-studied case of the RN black hole, there are scalar ZDMs which reduce to zero decay in the extremal limit $Q \to M$.
In this section, we examine the separated, decoupled GEM equations in the NERN background, and show that ZDMs exist for these perturbations as well.

The ZDMs of RN are purely decaying, like the $m=0$ ZDM modes of Kerr and the DF equation, and so they do not fit with the usual intuition into the nature of QNMs. 
Purely decaying perturbations of Schwarzschild have been discussed by Price~\cite{Price:1971fb,Price:1972pw,wheeler1972magic}, although the connection between this exponential decay and quasinormal modes remains unclear.
Purely decaying modes in RN have been described in~\cite{Andersson:1996xw,Andersson:2003fh}, and include the algebraically special modes~\cite{Chandra:1984a}, but to our knowledge none of these exhibit the slow decay rate we find, despite a large literature exploring the QNMs of extremal and nearly extremal RN black holes~\cite{LeaverRN,Onozawa:1995vu,Berti2009}.
Before exploring the existence of ZDMs for the NERN black hole, we review the fundamental equations for the perturbations of this spacetime.

\subsection{Perturbations of Reissner-Nordstr\"{o}m}

The problem of GEM perturbations for the RN spacetime closely parallels the investigation of perturbations of Schwarzschild using the Regge-Wheeler-Zerilli equations. 
The equations come in two sets, according to the parity of the perturbations, and it is known that the QNM spectrum of both sets is the same~\cite{ChandraBook,Dias:2015wqa}. 
This means that we can focus on the magnetic-parity perturbations [those which are multiplied by $(-1)^{l+1}$ under a parity transform]\footnote{We follow the convention of~\cite{Zerilli:1970wd}. Other studies refer to these modes as odd-parity and even-parity, or axial and polar, see e.g.~\cite{Pani:2013ija,Pani:2013wsa}.}, which gives two equations indexed by $j,k = 1,2$:
\begin{align}
\label{eq:RNwave}
\frac{d^2 Z_j}{dr_*^2}& + (\omega^2 - V_j) Z_j = 0 \,, \\
\label{eq:RNpot}
V_j & = \frac{\Delta}{r^5}\left[ l(l+1) r - q_k +\frac{4 Q^2}{r} \right] \,,
\end{align}
where $q_k$, $k \neq j$ indicates that $q_2\, (q_1)$ be used for $Z_1\, (Z_2)$, and
\begin{align}
q_1 &= 3 M  + \sqrt{9 M^2 +4 Q^2 [l(l+1)-2]} \,, \\
q_2 & = 6M - q_1 \,.
\end{align}
When $Q \to 0$, $Z_1$ obeys the equations for  magnetic-parity electromagnetic perturbations and $Z_2$ obeys the Regge-Wheeler equation for gravitational perturbations. In the above equations, $r_*$ and $\Delta$ are defined in the same way as in the KN spacetime, in the limit $a \to 0$. In particular, $\Delta$ has two roots which give the coordinate positions of the outer and inner event horizons, $r_\pm = M \pm M \sqrt{1 - Q^2/M^2}$.
We are interested in the nearly extremal limit, where $\sigma = (r_+ - r_-)/r_+ \ll 1$. We maintain the same notation as in Sec.~\ref{sec:DF}, which highlights many parallels between two analyses.

To search for ZDMs in the NERN spacetime analytically, we repeat the steps of the matched asymptotic expansion used for the DF equation in Sec.~\ref{sec:DF}. First we discuss the inner solution.

\subsection{The inner solution}

In the near horizon, nearly extremal limit [$x = (r-r_+)/r_+\ll 1$ and $\sigma \ll 1$, but without assuming that $\sigma/x$ is small], Eqs.~\eqref{eq:RNwave} and~\eqref{eq:RNpot} reduce at leading order to 
\begin{align}
y^2 & Z_j''(y) + y Z_j'(y) + V^y_j Z_j (y)= 0\,, \\
V^y_j & = \left(\frac{\hat \omega}{\sigma}\right)^2 +\frac{y[q_k /M -4 - l(l+1)]}{(1-y)^2} \,.
\end{align}
We recall the variable $y$ used in Sec.~\ref{sec:DF}, 
\begin{align}
y = \exp \left( \frac{\sigma r_*} {r_+} \right) \approx \frac{x}{x+\sigma} \,,
\end{align}
By making the replacements $  \hat \omega/\sigma= \varpi / 2 $ and
\begin{align}
 q_k /M -4 - l(l+1) & = \delta_j^2 + 1/4 \,, 
 \end{align}
we see that the near-horizon approximation for $Z_j$ reduces to the same equations as in the DF analysis, Eqs.~\eqref{eq:NHdiffeq} and~\eqref{eq:Ypot}, with $\zeta = 0$, $s = 0$, and 
\begin{align}
\delta_j & = i \left [L + (-1)^{j-1} \right]\,, &&  j  = 1,2\,,
\end{align}
recalling that $L = l+ 1/2$.
Note however that the form of $\delta_j$ differs slightly from the factor $\delta$ for the DF equation in the $a \to 0$ limit. 
In that case, $\delta = i L$ (and is independent of the spin $s$ of the test scalar field). 
For RN, $\delta_j$ is purely imaginary, and we take as our convention that $\delta_j$ is a positive imaginary number.
Again selecting the solution to Eq.~\eqref{eq:NHdiffeq} which has no waves emerging from the horizon and normalizing the amplitude of the solution to unity at the horizon, we have
\begin{align}
\label{eq:RNnearsln}
Z_j & = y^{- i \varpi} (1 - y)^{1/2 + i \delta_j} \, {}_2 F_1\left(\alpha,\beta,\gamma, y \right) \,, \\
\alpha & = 1/2 - i \varpi + i \delta_j \,, \qquad \beta = 1/2 + i \delta_j, \qquad \gamma = 1 - i\varpi \,.
\end{align}
As with the DF equation, we next wish to match this solution onto a solution in the outer region, where $x \gg \sigma$.

\subsection{Ansatz for matching}

We turn to the approximation of Eq.~\eqref{eq:RNwave} when we can take $(x + \sigma) \approx x$. 
Substituting in our definitions, we have after some manipulation,
\begin{align}
&x^2 \frac{d^2 Z_j}{dx^2} + \frac{2x}{(x+1)} \frac{d Z_j}{dx} \notag \\ & + \left[\frac{(x+1)^4}{x^2}  \omega^2 - l(l+1) + \frac{q_k}{x+1} + \frac{4}{(x+1)^2}\right] Z_j  = 0 \,,
\end{align}
We have not yet found a simple analytic solution that gives a convenient matching condition for small $x$. Using transformations such as $Z_j = \Delta^{s/2} r R$ yields promising forms of the equation for various choices of $s$, but none which allow for a straightforward matching analysis. 

Instead, motivated by past experience, we make an ansatz to complete the matching. By expanding the inner solution as in Sec.~\ref{sec:Matching} and Appendix~\ref{sec:MatchingApp}, we have
\begin{align}
\label{eq:RNmatch}
Z_j \to &\frac{\Gamma(2 i \delta_j) \Gamma(1-i \varpi)}{\Gamma(1/2 + i \delta_j)\Gamma(1/2 - i \varpi + i \delta_j)} \left( \frac{\sigma}{x} \right)^{1/2-i\delta_j}\notag \\
&+ (\delta_j \to - \delta_j) \,,
\end{align}
In the case of the DF equation in NEKN, the ZDM solutions correspond to the near vanishing of one of the two coefficients of $(\sigma/x)^{1/2 \pm i \delta}$ in the above expansion. This occurs at the zeros of one of the $1/\Gamma(w)$ factors, since $\Gamma(w)$ has poles at the negative integers. We make the ansatz that the corresponding Gamma function is also near its pole in RN. Investigating Eq.~\eqref{eq:RNmatch}, it is apparent that for the convention where $\delta_j$ is a positive imaginary number, the only possibility for fulfilling this criteria here is by taking 
\begin{align}
\label{eq:RNprematch}
1/2 - i \varpi - i \delta_j = - n \,,
\end{align}
which gives 
\begin{align}
\label{eq:RNfreq}
\omega = -i \frac{\sigma}{2r_+} \left[ |\delta_j| + \left(n + \frac 12 \right)\right] \,.
\end{align}
The fact that $\delta_j$ is pure imaginary indicates the presence of DMs in addition to the ZDMs with frequencies given by Eq.~\eqref{eq:RNfreq}, which is of course what is observed. 
These DMs are the usual QNMs of extremal and nearly extremal RN which have been the subject of past study.

We can provide a heuristic argument in support of our ansatz. 
When $x \gg \sigma$ and $x \lesssim 1$, the two terms in the wave function~\eqref{eq:RNmatch} have distinctly different behavior while approaching the edge of the near-horizon region (corresponding to a decreasing $\sigma/x$). 
In the first term, displayed explicitly in Eq.~\eqref{eq:RNmatch}, $\sigma /x$ has an exponent  $1/2 + |\delta_j| >0$ and so is decaying as $x$ increases. In the second term, where the first term has the replacement $\delta_j \to - \delta_j$, the exponent of $\sigma/x$ is negative and so the term is growing. 
Our matching condition sets this growing term to zero, which is reasonable: when $x \sim 1$, the growing term will have a size $\sim (1/\sigma)^{|\delta_j| -1/2} \gg 1$. 
In the DF case, if the amplitude of this term is not suppressed, this term is too large to match onto the outer solution. 
It seems reasonable that in general an outer solution, which is regular as $\sigma \to 0$, cannot match onto this growing term if the amplitude is not suppressed.
Unless the perturbations have nearly zero amplitude in the near-horizon region, we would encounter unnaturally strong perturbations in the matching region, invalidating the approximate formalism even for QNMs sourced by small initial data.
At the very least, a large amplitude in the matching region goes against the intuition that the ZDMs are concentrated or trapped near the horizon \cite{Andersson2000}. 
This possibility can be avoided for frequencies which eliminate this troublesome growing term.

Note that if an outer solution were available to us, we would find that Eq.~\eqref{eq:RNfreq} is corrected by a small $\eta$ as in the case of the ZDM frequencies predicted by the DF matching analysis of Sec~\ref{sec:DF}. 
This correction would mean that the second term of Eq.~\eqref{eq:RNmatch} is not precisely zero, but its large size is compensated by the small amplitude factor. 
The balance of these two effects would allow for a matching onto the outer solution to take place, as is done for the DF equation.

In Sec.~\ref{sec:RNNumerics} we show that Eq.~\eqref{eq:RNfreq} gives the correct decay rates for the RN ZDMs. 
The fact that this equation differs from the $a \to 0$ limit of the DF prediction, Eq.~\eqref{eq:DFfreq} (recalling that $\delta_j \neq \delta$) shows that the DF equation fails to describe the QNMs of the nearly extremal RN spacetime outside of the scalar case. 
This is in contrast to the situation where the charge of the NEKN spacetime is small, where the DF equation provides the leading frequency corrections to the Kerr ZDMs~\cite{Mark:2014aja}, and shows that we cannot hope that the DF equation is accurate for all NEKN black holes.

\subsection{WKB analysis}

For completeness, we include the WKB analysis of Eqs.~\eqref{eq:RNwave} and~\eqref{eq:RNpot}. The form of Eq.~\eqref{eq:RNwave} allows us to immediately use the methods of~\cite{Schutz:1985zz,IyerWill1987}. We recall our definitions $L = l + 1/2 \gg 1$, $\Omega_R = \omega_R/L$, and in the notation of~\cite{Schutz:1985zz} we define
\begin{align}
\mathcal Q = \omega^2 - V_j \approx L^2 \left(\Omega_R^2 - \frac{\Delta}{r^4} \right) \,.
\end{align}
The conditions that $\mathcal Q = 0$ and $d \mathcal Q/dr_* = 0$ at the WKB frequency identifies the peak $r_0$ and gives $\Omega_R$,
\begin{align}
r_0 & = \frac 12 \left( 3 M + \sqrt{9 M^2 - 8 Q^2}\right) \,, &
\Omega_R & = \left. \frac{\sqrt{\Delta}}{r^2} \right|_{r_0} \,,
\end{align}
The curvature at the extrema of the potential determines the decay rate of the mode, Eq.~\eqref{eq:WKBOmegaI}. The result for RN is
\begin{align}
\Omega_I = \frac{\sqrt{3 M^2 r_0 - Q^2 (M+2r_0)}}{r_0^{5/2}}  \,.
\end{align}
These results agree with the literature~\cite{Kokkotas:1988fm,Andersson:1996xw,Mashhoon1985,Cardoso2009}. In the extremal limit,
\begin{align}
r_0 &= 2 M \,, & \Omega_R & = \frac{1}{4M} \,, & \Omega_I & = \frac{1}{4M \sqrt{2}}\,. 
\end{align}
In the context of this study, the important point is that the decay rate remains nonzero for all $Q$, and so this analysis can approximate only the usual DMs of RN in the extremal limit. It does not access the ZDMs.

\subsection{Numerical results}
\label{sec:RNNumerics}

\begin{figure}[tb]
\includegraphics[width =.95 \columnwidth]{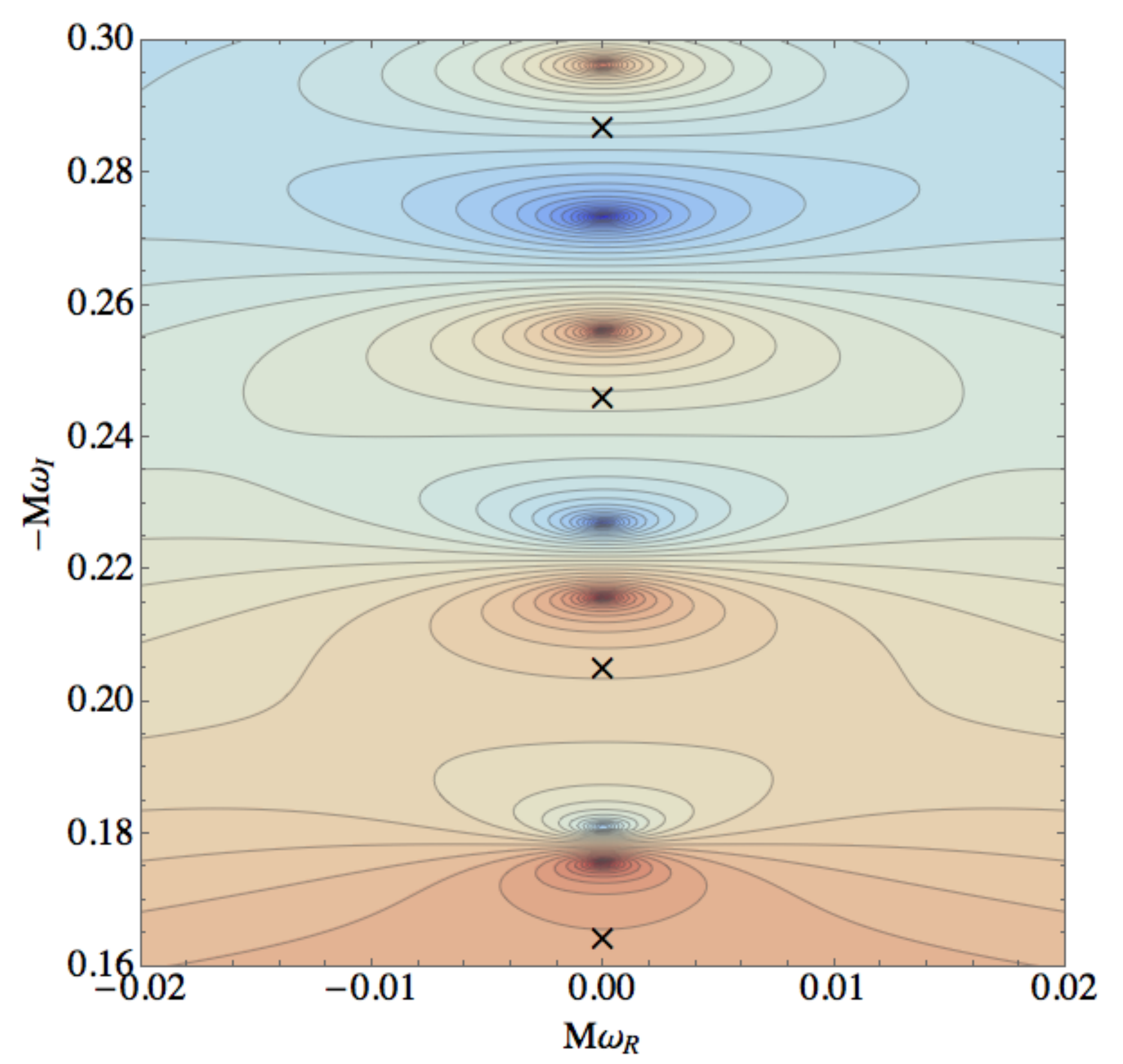}
\includegraphics[width =.95 \columnwidth]{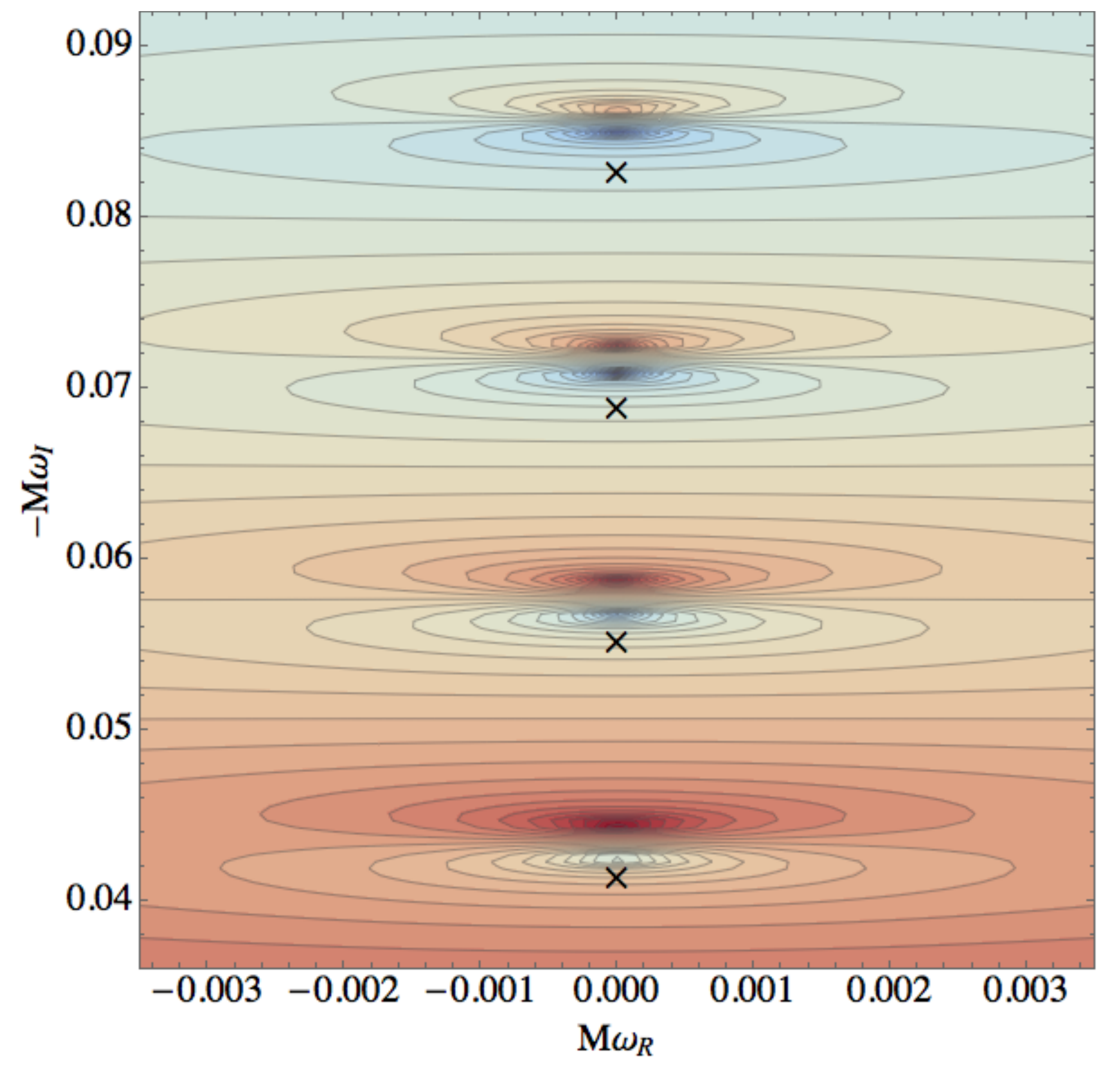}
\caption{Contour plots of the logarithm of the magnitude of Leaver's RN continued fraction, illustrating the $l=1$ ``electromagnetic'' ZDMs (as defined in the limit $Q \to 0$) of NERN. The black crosses are the predictions of Eq.~\eqref{eq:RNfreq}. The plots focus on the four modes with the smallest decay rates. {\it Top panel}: The case $Q=0.999M$.  {\it Bottom panel}: The more extreme case $Q=0.9999M$. }
\label{fig:RNClose}
\end{figure}

While we consider the ansatz for the outer solution to be well motivated, we must numerically check the expression for the ZDM frequencies of NERN, Eq.~\eqref{eq:RNfreq}. 
In this section, we show that Eq.~\eqref{eq:RNfreq} is accurate using the methods of Sec.~\ref{sec:DFNumerics}, namely via contour plots of the logarithm of the continued fraction $\mathcal C^r$ and numerical calculations of the residual error $\Delta \omega$. 
We find that residual error in Eq.~\eqref{eq:RNfreq} scales identically to the residual error in the DF ZDM formula; $\Delta \omega \sim O[(n+1/2)\sigma^{2}]$. 
Together with the WKB results for the damped modes, this analysis demonstrates that the RN QNM spectrum also undergoes a bifurcation as $\sigma \to 0$ .

To numerically compute the QNM frequencies, we once again use Leaver's method, which can be extended with some effort to RN black holes~\cite{LeaverRN}. For RN, the angular problem decouples from the radial problem and is solved by scalar, vector, and tensor spherical harmonics with known eigenvalues.  The radial wave functions can be expanded as a power series whose coefficients obey a four-term recurrence relation. Through Gaussian elimination, these can be converted into three-term recurrence relations and then into a radial continued fraction $\mathcal C^r$ whose roots give the QNM frequencies.

In Fig.~\ref{fig:RNClose}, we present two contour plots of the logarithm of $|\mathcal C^r|$ in the complex-$\omega$ plane, for the case $j=1$ (electromagnetic-type) and $l=1$, for nearly extremal values of charge $Q = 0.999M$ and $Q= 0.9999M$. These plots demonstrate the existence of $j=1$, $l=1$ ZDMs lying on the imaginary axis. Visually, we again see agreement with the prediction of Eq.~\eqref{eq:RNfreq} (black crosses). 

\begin{figure*}[tb]
\includegraphics[width =1.0 \columnwidth]{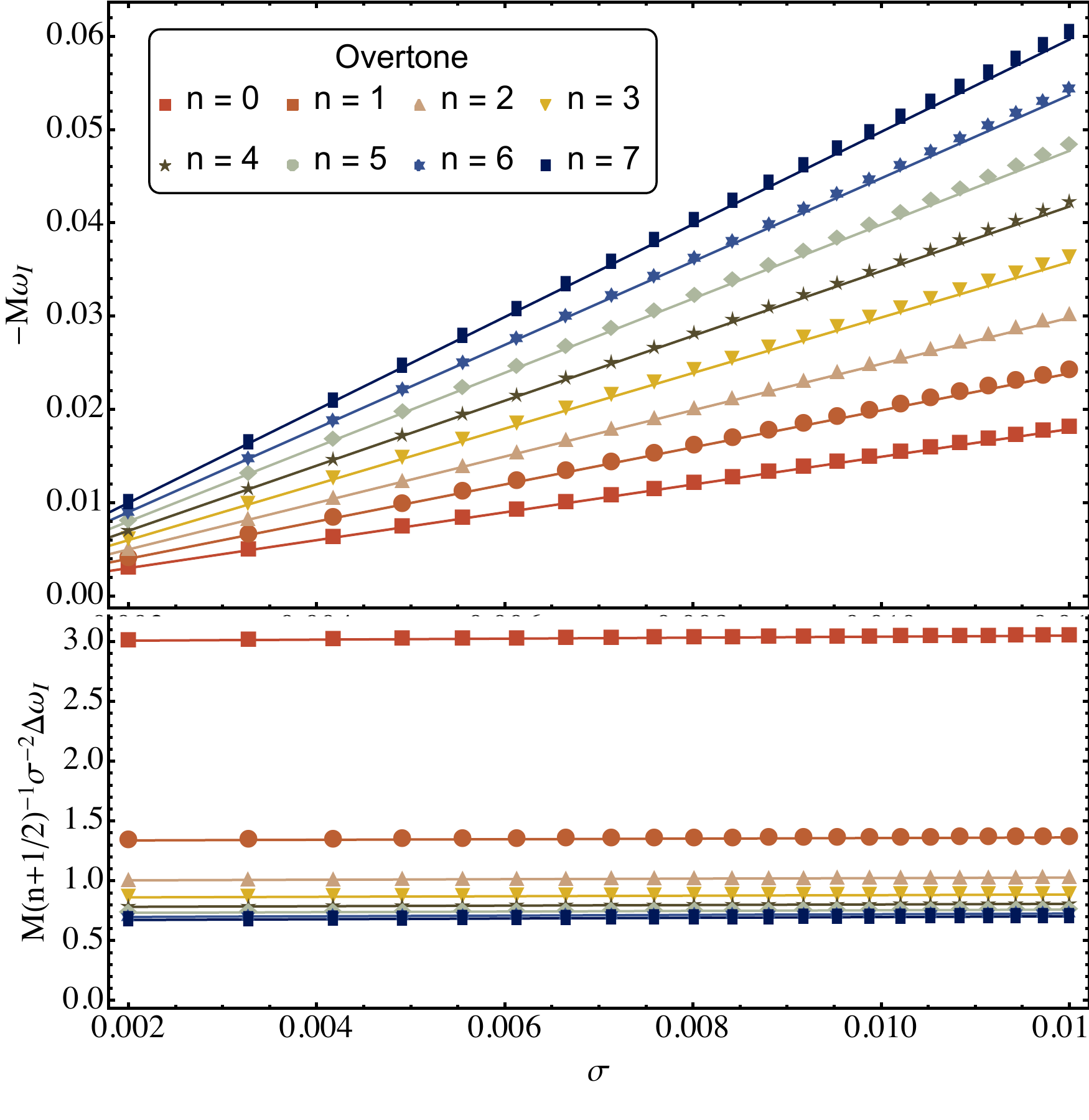}
\includegraphics[width =1.0 \columnwidth]{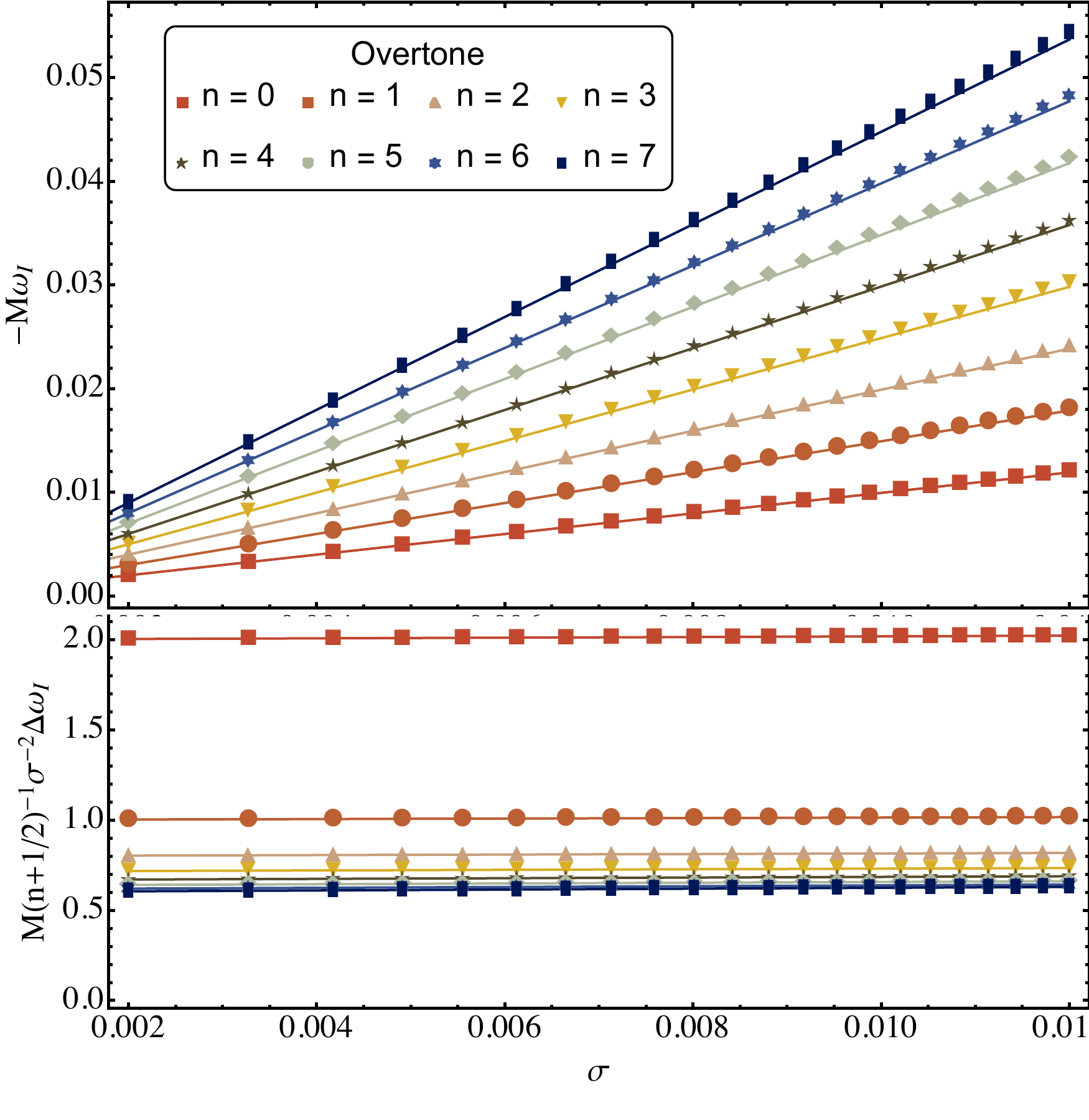} \\
\includegraphics[width =1.0 \columnwidth]{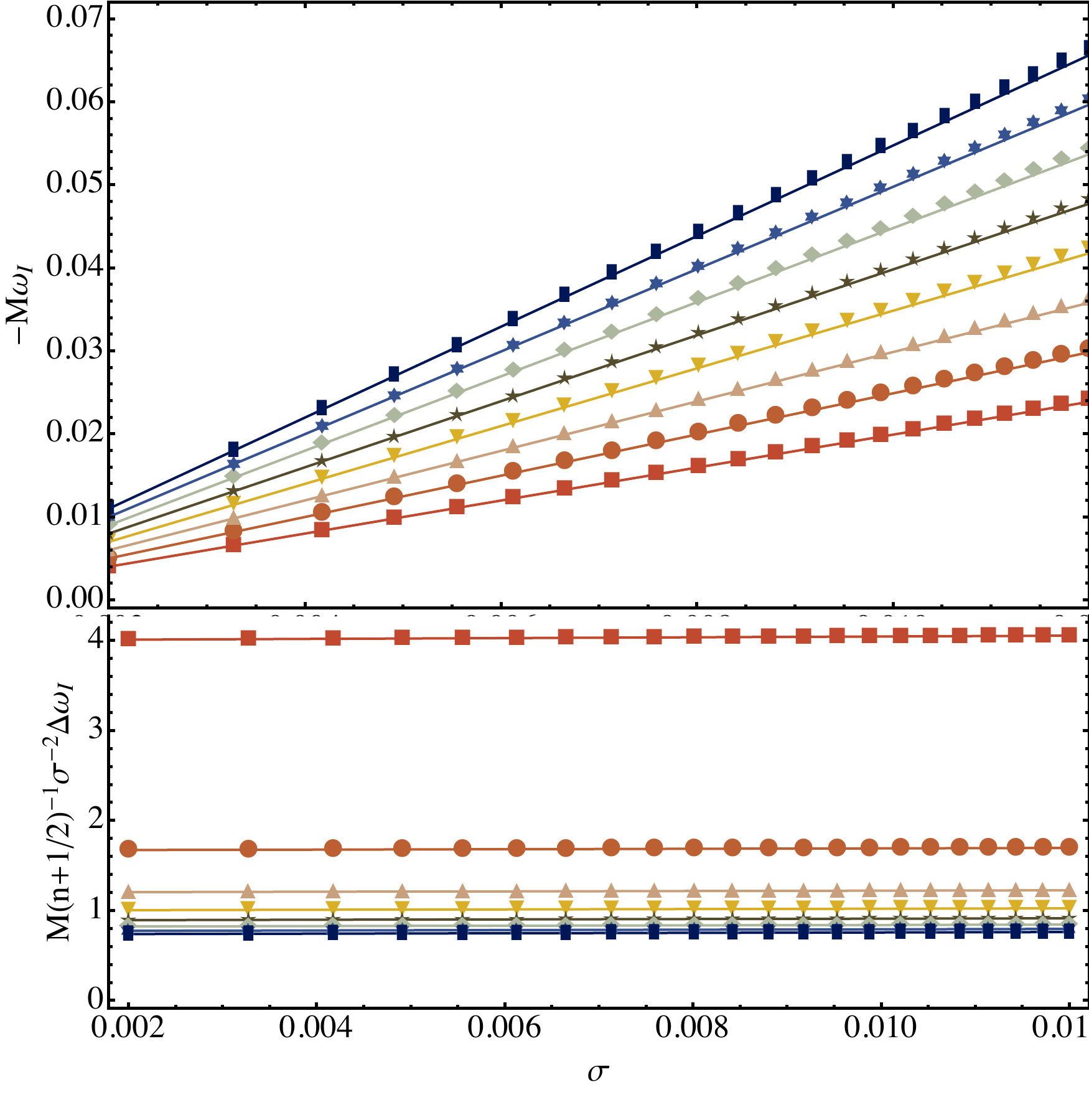} 
\includegraphics[width =1.0 \columnwidth]{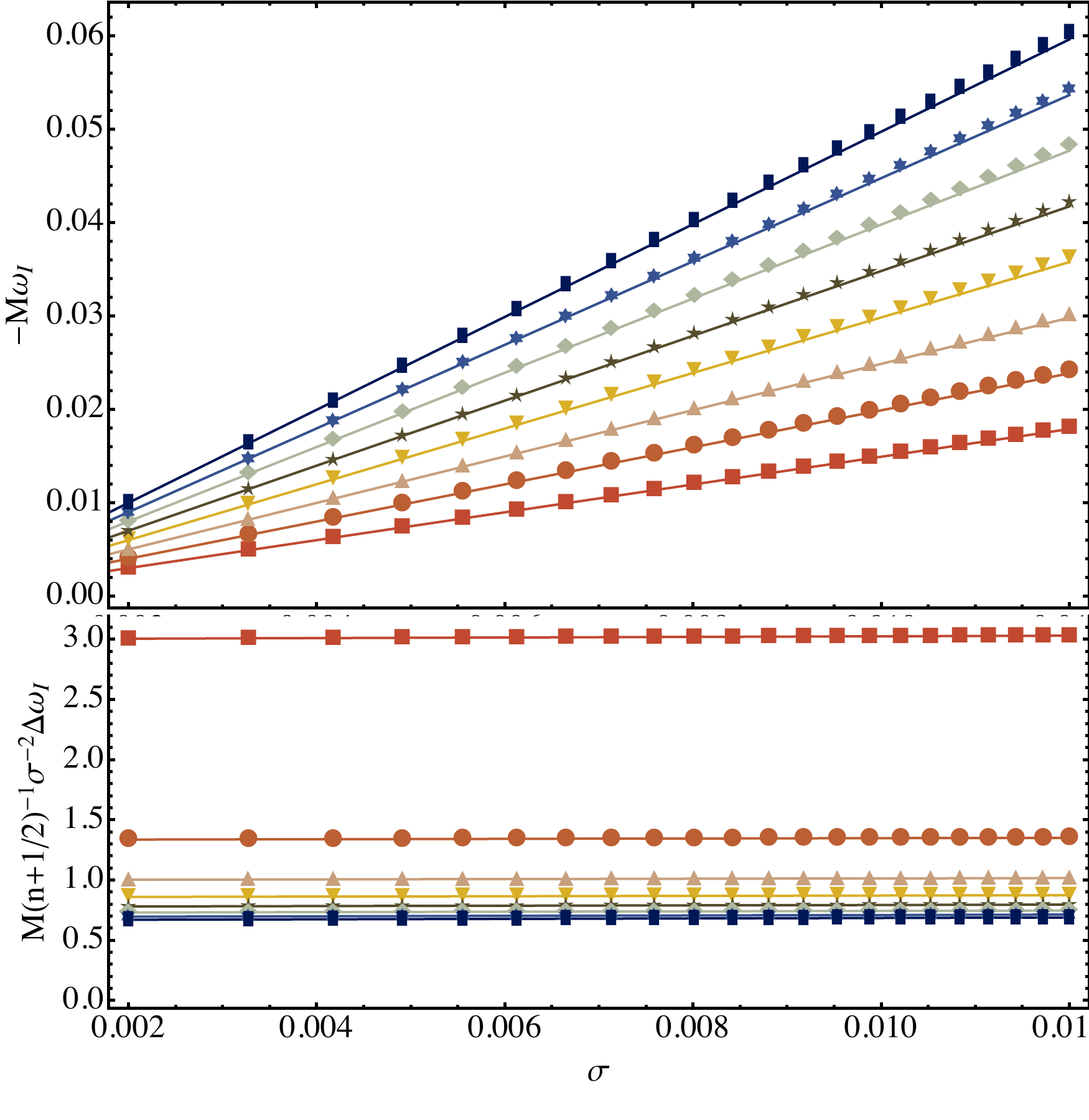}
\vspace{-5mm}
\caption{ZDM frequencies for perturbations of the RN black hole. The top of each panel plots the eight lowest overtones for several values of $\sigma$, calculated with Leaver's method (points). We also plot the analytical RN ZDM frequency predictions of Eq.~\eqref{eq:RNfreq} for these overtones (lines). The bottom of each panel plots the scaled residual error of Eq.~\eqref{eq:RNfreq}, $M\sigma^{-2}(n+1/2)^{-1}\Delta \omega$ versus $\sigma$, and the lines here simply join the points. 
Following each line from right to left demonstrates that the residual is $O(\sigma^2)$ and following the calculations downward at a fixed value of $\sigma$ demonstrates that the residual is $O(n+1/2)$ at large $n$. 
{\it Top left}: The case  $j=1$, $l=1$. {\it Bottom left}: The case $j =1$, $l = 2$. {\it Top right}: The case $j = 2$, $l=2$. {\it Bottom right}: The case $j = 2$, $l=3$.}
\label{fig:slRN}
\end{figure*}

Figure~\ref{fig:slRN} presents a broader and more quantitative analysis. Each panel corresponds to a different value of $l$ and $j$ (mode type). The top of each panel plots the lowest eight overtones of the ZDMs, found using Leaver's method, along with the predictions of Eq.~\eqref{eq:RNfreq}. In the bottom of each panel, we plot the scaled residual error $M(n+1/2)^{-1}\sigma^{-2}\Delta\omega$ versus $\sigma$. For each overtone, one can check the $\sigma^2$ scaling of the residual by following the corresponding line. For each value of $\sigma$, one can check the $(n+1/2)^{-1}$ scaling by following the calculations vertically downward. 
While these results do not represent a comprehensive search for the ZDMs of RN, they give us confidence that the matching ansatz gives the correct expression for these frequencies. Importantly, our analysis establishes the existence of ZDMs of the GEM perturbations of RN black holes for the first time to our 
knowledge\footnote{During the completion of this work we became aware of the study~\cite{Couch:2012zz}, which uses methods analogous to Leaver's method on the nearly extremal DF equation in the RN limit. That study identifies the scalar ZDMs, but incorrectly claims that the results apply to electromagnetic and gravitational perturbations.}.

\section{Conclusions}
\label{sec:Conclusions}

In this paper we have given an overview of the QNMs of nearly extremal Kerr-Newman black holes. 
While many of the results in Sec.~\ref{sec:DF} have appeared elsewhere, there are many contradictory results in the literature. We have reviewed the derivation of the ZDM frequencies for NEKN black holes, using matched asymptotic expansions. 
Using the WKB approximation for scalar fields in KN, we have discussed the existence of damped modes and given approximate formulas for these frequencies. 
Finally, using Leaver's method, we have validated these approximations, and effectively measured the higher order corrections to the nearly extremal and WKB approximations.

By carrying out this analysis using the DF equation for spin-weighted scalars, the results of Sec.~\ref{sec:DF} can be compared to results for the true GEM modes of NEKN, in order to see how this simplistic model performs, for example by a careful comparison to numerical results. 
This is left for future studies, although we reiterate that the DF equation correctly predicts the small-charge corrections to the ZDMs of nearly extremal Kerr black holes~\cite{Mark:2014aja}.

Since the case of scalar QNMs in NERN follows immediately from the results of Sec.~\ref{sec:DF}, in Sec.~\ref{sec:RN} we have investigated the coupled GEM equations of NERN. In this case, we have shown that ZDMs exist alongside the well known DMs, and given a frequency formula for these modes using a matching ansatz. A numerical study using Leaver's method confirms this ansatz and again the residual errors provide higher order corrections. 
The ZDM frequency formula differs from that of the spin-weighted scalars found using the DF equation, indicating that the DF equation cannot accurately describe the ZDM frequencies for all spinning, charged black holes. 
For completeness, we have provided the WKB formulas for RN, and examined its extremal limit, concluding that the technique only describes damped modes.

While this work demonstrates the existence of a family of purely damped QNMs for RN, it is unclear what the implications of these modes are. They may assist in the shedding of black hole hair following collapse to an RN black hole, as is the case for exponentially decaying modes in Schwarzschild~\cite{Price:1971fb,Price:1972pw,wheeler1972magic}. A careful analysis of the excitation of QNMs would be required to assess the importance of these modes and their physical meaning.

Looking ahead, the daunting problem of gaining an analytic understanding of coupled GEM equations in the nearly extremal case remains.
Two lines of evidence indicate that the GEM perturbations of NEKN admit ZDMs.
The first is the weakly charged, rapidly rotating case discussed in Mark {\it et al.}~\cite{Mark:2014aja}.
That study computed the QNM frequencies of weakly charged Kerr black holes in the form $\omega \approx \omega^{(0)} + Q^2 \omega^{(1)}$, where $\omega^{(0)}$ is the Kerr value. Mark {\it et al.}~showed that the DF equation provides a complete accounting of the frequency corrections $\omega^{(1)}$ to the gravitational and electromagnetic ZDMs in Kerr as the black hole angular momentum increases towards extremality.
The coefficients $\omega^{(1)}$ also begin to diverge in this limit, although they are  controlled by the smallness of $Q/Q_{\rm max}$, where $Q_{\rm max}$ is the charge of the extremal KN black hole at a given $a$.
Both of these behaviors can be explained if the full KN QNMs are ZDMs, whose frequencies are $m\Omega_H$ at leading order, with corrections proportional to the surface gravity $\kappa$.
Reexpanding such frequencies in small charge compared to $Q_{\rm max}$ recovers the apparent divergences of $\omega^{(1)}$ seen in that study, naturally suppressing them by $Q/Q_{\rm max}$.
Meanwhile the increasing accuracy of the DF equation can be understood by examining the near-horizon, near-extremal scalings of the ZDM wave functions of Kerr \cite{Mark:2014aja}, although the reason for these scalings remains a puzzle.

The other, even more compelling line of evidence is provided by the recent numerical investigations of the QNMs of KN.
The numerical results of~\cite{Dias:2015wqa} show definitively the existence of ZDMs in KN. In that study, a GEM mode with $l=m=2$ showed the behavior $\omega_R \sim m \Omega_H$ and $\omega_I \to 0$ in the extremal limit, for all values of $a$.
The fact that this occurs even when $Q \geq 0.5 M$ indicates that the search of~\cite{Dias:2015wqa} identifies the ZDMs, even in the regime where we expect DMs and where spectrum bifurcation might confuse a numerical search. 
Since~\cite{Dias:2015wqa} focused on only the lowest overtones (defined as having the smallest decay rate), future numerical studies will be required to understand the existence and behavior of the GEM damped modes of the KN black hole.

The dependence of the ZDM frequency~\eqref{eq:DFfreq} on spin and charge also explains the frequency behavior seen in the numerical simulations presented in~\cite{Zilhao:2014wqa}, as pointed out in a remark by Hod~\cite{Hod:2014uqa}. 
Those numerical simulations evolved perturbations of NEKN black holes using the full Einstein-Maxwell equations, and argued that for a range of values of $Q$ the perturbations had frequencies and decay rates dependent primarily on the combination $a/a_{\rm max} = a/ \sqrt{1 - Q^2/M^2}$. 
At first glance, this is in contradiction with our formula~\eqref{eq:DFfreq}. 
In fact the proposed relation gives almost the same frequencies as~\eqref{eq:DFfreq} for the cases $a\gtrsim0.9M, \, Q \lesssim 0.4 M$.
Only a precise analysis of the frequencies and decay rates, such as that given in \cite{Dias:2015wqa}, can differentiate the proposed relation from the one derived here.
Hod also notes that in the case presented in \cite{Zilhao:2014wqa} where $Q$ is large and the KN black hole is nearly extremal, $\Omega_H$ gives a good accounting for the observed frequency of the oscillations.

The success of the nearly extremal approximation for describing ZDMs of scalar modes in RN and both scalar and GEM perturbations of RN begs the question of whether such methods can be applied to the full, coupled GEM perturbations of Kerr-Newman.
Unfortunately, a naive near-horizon, nearly extremal scaling analysis indicates that these equations remain coupled in this limit, and this coupling in turn obstructs separation of the differential equations.
Nevertheless, the results of this paper, many previous studies, and recent comprehensive numerical results~\cite{Zilhao:2014wqa,Dias:2015wqa} all indicate that the ZDMs of the full coupled perturbations of NEKN obey a simple frequency formula like Eq.~\eqref{eq:DFfreq}. 
The challenge is to show that this is so, and provide an analytic expression for the factor of $\delta(a)$.
The wealth of progress in studying perturbations of KN black holes in the past several years places this goal within reach.
It may be that the connection to conformal field theories available in the near-horizon region of NEKN~\cite{Bardeen:1999px,Guica2009,Hartman:2008pb,Hartman:2009nz,Porfyriadis:2014fja} will allow for the solution of this problem, or at least to further connections to quantum theories.
Another promising avenue is the application of WKB techniques to the coupled GEM equations.
In the WKB limit, the differences between the DF and full GEM predictions for RN vanish, although the equations describe very different kinds of perturbations. 
It may be that this fact carries through to the rotating KN black hole, in which case the DF WKB predictions would give an accurate accounting of the high-frequency GEM modes of KN.
We leave the investigation of this possibility to future studies.

\acknowledgements

We thank Emanuele Berti, Yanbei Chen, David Nichols, Huan Yang, An{\i}l Zengino\u{g}lu and Fan Zhang for previous collaboration and valuable discussion on the topic of the QNMs of nearly extremal black holes. 
We are especially grateful to Huan Yang and Yanbei Chen for collaboration on additional studies of Kerr-Newman black holes and the WKB approximation for QNMs, which this study grew out of, and to Yanbei Chen and Fan Zhang for past collaboration on the implementation of Leaver's method in the Kerr spacetime. 
We also thank Sam Gralla and Alexandru Lupsasca for insights into the near-horizon region of nearly extremal Kerr. 
A.~Z.~was supported by the Beatrice and Vincent Tremaine Postdoctoral Fellowship at the Canadian Institute for Theoretical Astrophysics during much of this work. 
Z.~M.~is supported by NSF Grant No. PHY-1404569,
CAREER Grant No. 0956189, the David and Barbara Groce Startup Fund at Caltech, and the Brinson Foundation. 
A portion of this work was carried out at the Perimeter Institute for Theoretical Physics. 
Research at Perimeter Institute is supported by the government of Canada and by the Province of Ontario though Ministry of Research and Innovation.

\appendix
\begin{widetext}

\section{Details of the matching calculation}
\label{sec:MatchingApp}

In this Appendix we provide some supplementary equations related to the matching analysis of Sec.~\ref{sec:Matching}.
First we consider the outer solution, Eq.~\eqref{eq:OuterSln}.
In the limit $x \to \infty$, we use the expansion~\cite{nist}
\begin{align}
{}_1 F_1(a, b, 2 i \hat \omega x) \to \frac{\Gamma[b]}{\Gamma[a]} e^{2 i \hat \omega x}(2 i \hat \omega x)^{a-b} + \frac{\Gamma[b]}{\Gamma[b -a ]} (-2 i \hat \omega x)^{-a} \,. 
\end{align}
The first term above contributes to the part of the radial function which behaves as $R \propto e^{i \omega r_*}$ and so is an outgoing solution. Similarly, the second term contributes to the ingoing solution, which can be eliminated by a particular choice of $A$ and $B$.
The requirement that we have only outgoing waves provides the condition
\begin{align}
\label{eq:Ratio1}
\mathcal R = \frac{A}{B} & = \frac{\Gamma[-2 i \delta] \Gamma[1/2 + s + i \delta - 2 i \hat \omega]}{\Gamma[2  i \delta] \Gamma[1/2 + s - i \delta - 2 i \hat \omega]} e^{\pi \delta+ 2 i \delta \log 2 \hat \omega} \,.
\end{align}
We can also identify the outgoing and ingoing wave amplitudes in the general scattering problem. We have
\begin{align}
\label{eq:OuterAmp1}
\frac{Z^{\rm out}}{Z^{\rm hole}} & = A (2 i \hat \omega)^{-1/2 -s - i \delta + 2 i \hat \omega} \frac{\Gamma[1 + 2 i \delta]}{\Gamma[1/2 - s + i \delta + 2 i \hat \omega]} +B (\delta \to \delta) \,, \\
\label{eq:OuterAmp2}
\frac{Z^{\rm in}}{Z^{\rm hole}} & = A (-2 i \hat \omega)^{-1/2 + s - i \delta - 2 i \hat \omega} \frac{\Gamma[1 + 2 i \delta]}{\Gamma[1/2 + s + i \delta - 2 i \hat \omega]} + B (\delta \to - \delta)\,.
\end{align}
Here we have normalized each amplitude by the amplitude of the wave function at the horizon, $Z^{\rm hole}$. Elsewhere in this paper, we have assumed that $Z^{\rm hole} = 1$.

Meanwhile, in the limit of small $x$, the outer solution takes the simple form
\begin{align}
R = A x^{-1/2 - s + i \delta} + B x^{-1/2 -s - i \delta} \,.
\end{align}
The inner region provides a second condition by matching this to the inner solution. For this we transform the domain of the hypergeometric function  ${}_2 F_1$ by taking $z = 1-y$ and using the identity~\cite{nist}
\begin{align}
\label{eq:Inversion}
{}_2 F_1( \alpha, \beta, \gamma, y) & = \frac{\Gamma[\gamma] \Gamma[-2 i \delta]}{\Gamma[\gamma- \alpha]\Gamma[\gamma-\beta]} \,{}_2 F_1(\alpha, \beta, 1+ 2 i \delta, z) + \frac{\Gamma[\gamma]\Gamma[2 i \delta]}{\Gamma[\alpha]\Gamma[\beta]} z^{-2 i \delta} \,{}_2 F_1(\gamma - \alpha, \gamma -\beta, 1 -2 i \delta, z) \,.
\end{align}
It is important that in KN the parameters of the hypergeometric function obey $\gamma - \alpha - \beta = - 2 i \delta$ just as in Kerr, which allows the matching to occur. This has been used to simplify some of the terms in the above identity.
Note also that $\gamma - \alpha  = \beta|_{\delta \to -\delta}$ and $ \gamma - \beta = \alpha|_{\delta \to -\delta}$, demonstrating the symmetry of these equations under the change of the sign of $\delta$. 
This means that we can use the convention that $\delta$ is a positive real or imaginary number without loss of generality.

Next we take the limit $z \to 0$, which sets the hypergeometric functions in Eq.~\eqref{eq:Inversion} to unity.
Some useful identities for the matching are
\begin{align}
R  \approx \frac{r_+^{-s} \, x^{-s}}{\sqrt{r_+^2 +a^2}} u \,, && z \approx \frac{\sigma}{x} \,,
\end{align}
and we recall that $u = y^{-p} (1-y)^{-q} {}_2 F_1(\alpha, \beta, \gamma, y)$. 
Combining all of these equations gives us an expression for the inner solution,
\begin{align}
\label{eq:innmatch}
R & \approx \frac{r_+^{-s}}{\sqrt{r_+^2 +a^2}} x^{-s} \left(\frac{\sigma}{x} \right)^{1/2+i\delta} \left [\frac{\Gamma[\gamma] \Gamma[-2 i \delta]}{\Gamma[\gamma- \alpha]\Gamma[\gamma-\beta]} + \frac{\Gamma[\gamma]\Gamma[2 i \delta]}{\Gamma[\alpha]\Gamma[\beta]}\left( \frac{\sigma}{x}\right)^{-2 i \delta} \right] \,,
\end{align}

The matching gives 
\begin{align}
\label{eq:InnerAmp}
A & = \frac{r_+^{-s}\sigma^{1/2 - i \delta}}{\sqrt{r_+^2 + a^2}} \frac{\Gamma[\gamma] \Gamma[2 i \delta]}{\Gamma[\alpha] \Gamma[\beta]}\,, &
B & = \frac{r_+^{-s}\sigma^{1/2 + i \delta}}{\sqrt{r_+^2 + a^2}} \frac{\Gamma[\gamma] \Gamma[-2i\delta]}{\Gamma[\gamma - \alpha] \Gamma[\gamma - \beta]} \,,
\end{align}
so that 
\begin{align}
\label{eq:Ratio2}
\frac{A}{B} & = \sigma^{- 2 i \delta} \frac{\Gamma[2i\delta]}{\Gamma[-2 i \delta]}\frac{\Gamma[\gamma - \alpha]\Gamma[\gamma - \beta]}{\Gamma[\alpha]\Gamma[\beta]} \,.
\end{align}
In the general scattering problem, Eqs.~\eqref{eq:OuterAmp1}, \eqref{eq:OuterAmp2}, and \eqref{eq:InnerAmp} give the full expressions for the amplitudes, or equivalently the reflection and transmission coefficients of scalar waves in NEKN. 
The scattering amplitudes given here also allow for the calculation of QNM excitation factors, following the steps in~\cite{Yang:2013uba}. 
Since we do not deal with excitation of QNMs here, we omit these lengthy expressions.

The conditions from Eqs.~\eqref{eq:Ratio1} and~\eqref{eq:Ratio2} can be satisfied if one of the Gamma functions in~\eqref{eq:Ratio2} is near to a pole at the negative integers. 
In the case where $\delta$ is pure imaginary with our convention, Eq.~\eqref{eq:Ratio2} is suppressed by the smallness of $\sigma$, whereas Eq.~\eqref{eq:Ratio1} has no sensitive dependence on $\sigma$. 
This difference in behavior can be compensated by having one of the Gamma functions take a large value, i.e.\ be near its poles.
Meanwhile, when $\delta$ is real, then Eq.~\eqref{eq:Ratio1} exhibits rapid oscillation in phase when $\sigma$ is changed by a small amount; there is no such sensitive phase dependence on $\sigma$ in $\mathcal R$. 
The same assumption, that one of the Gamma functions is near its pole, can be used to compensate for this phase dependence, if the shift of the argument of the Gamma from its pole absorbs this rapid phase variation.
These considerations motivate the condition $\Gamma[\gamma - \beta] = \Gamma[-n - i \eta]$. Since $\gamma - \beta$ depends on the frequency through $\varpi$, this condition selects a particular QNM frequency. 
Expanding this condition in small $\eta$, we have $1/\Gamma[-n - i \eta] = (-1)^n (n!) (-i \eta) + O(\eta^2)$. We can solve the matching condition for $\eta$, giving an expression for the correction to $\omega$,
\begin{align}
\label{eq:EtaEqn}
\eta & = \mathcal R^{-1} (-1)^n \frac{i  \sigma^{- 2 i \delta}}{n!}  \frac{\Gamma[2i\delta]}{\Gamma[-2 i \delta]}\frac{\Gamma[\gamma - \alpha]}{\Gamma[\alpha]\Gamma[\beta]}\,.
\end{align}
As discussed in~\cite{Yang:2013uba}, $\eta$ is generally quite small, with the exception of cases where $\delta^2$ is small and negative, in which case it can be order unity or greater. 
Although the possibility has not been explored in detail, there may even be situations where $\eta$ could be large enough to invalidate the approximation used above to find a closed form solution for the ZDM frequencies.
We do not incorporate the correction $\eta$ into our frequency formula in this paper, but in cases where $\eta$ is a significant contribution to the $O(\sigma)$ frequency corrections, Eq.~\eqref{eq:EtaEqn} can be used to augment the ZDM frequency formula. 
In addition, whenever $\eta \gtrsim \sigma$, it dominates the residual error in our frequency formula. 
In Sec.~\ref{sec:DFNumerics} for $l=m=2$, $s=1$ and $\delta^2>0$, we find that the term $\sigma \eta$ prevents $O(\sigma^2)$ scaling of the residual errors at small values of $\sigma$.

\section{The WKB analysis of the Dudley-Finley equations}
\label{sec:WKBApp}

\begin{figure*}[t]
\includegraphics[width = 0.48 \columnwidth]{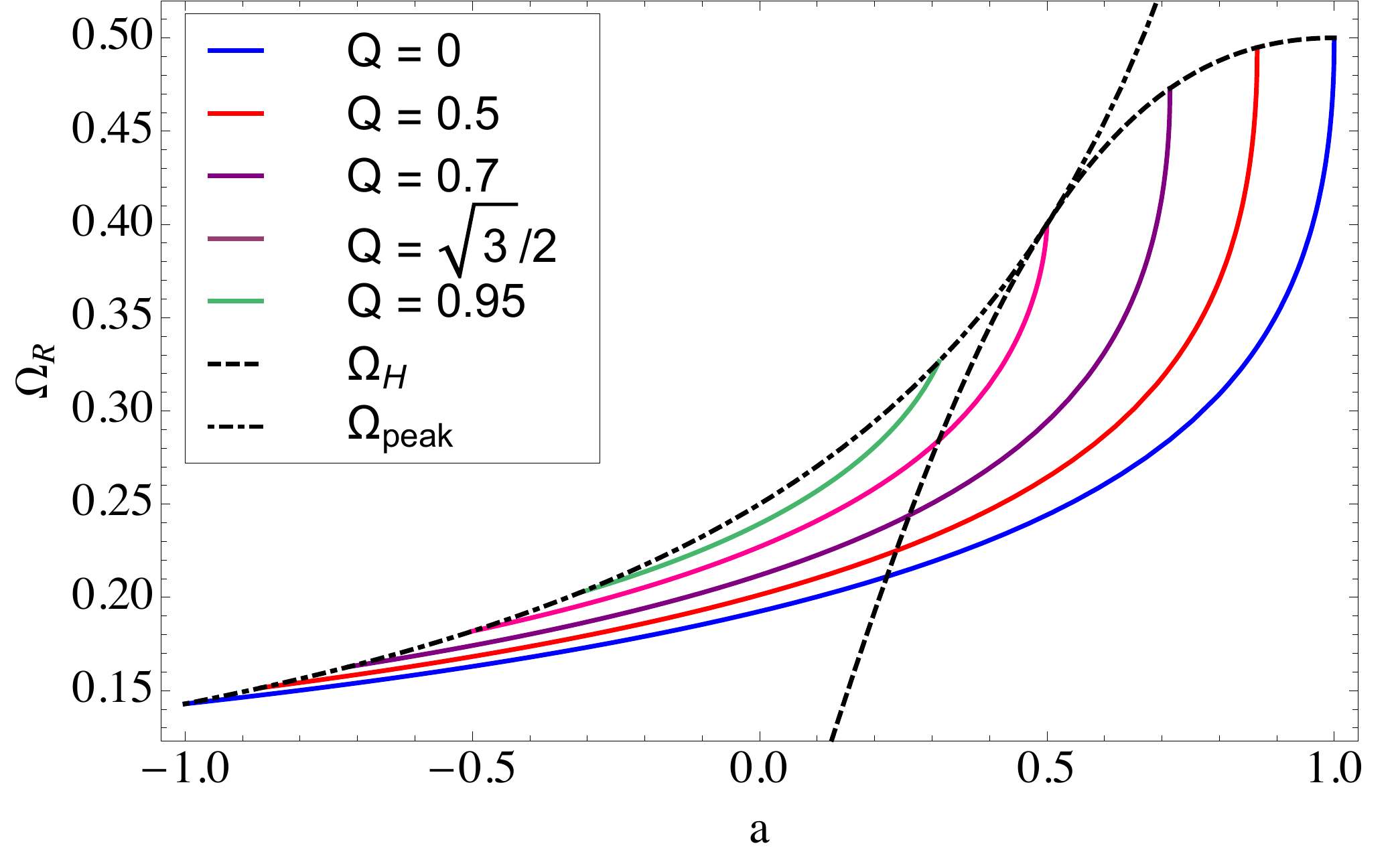}
\includegraphics[width = 0.48 \columnwidth]{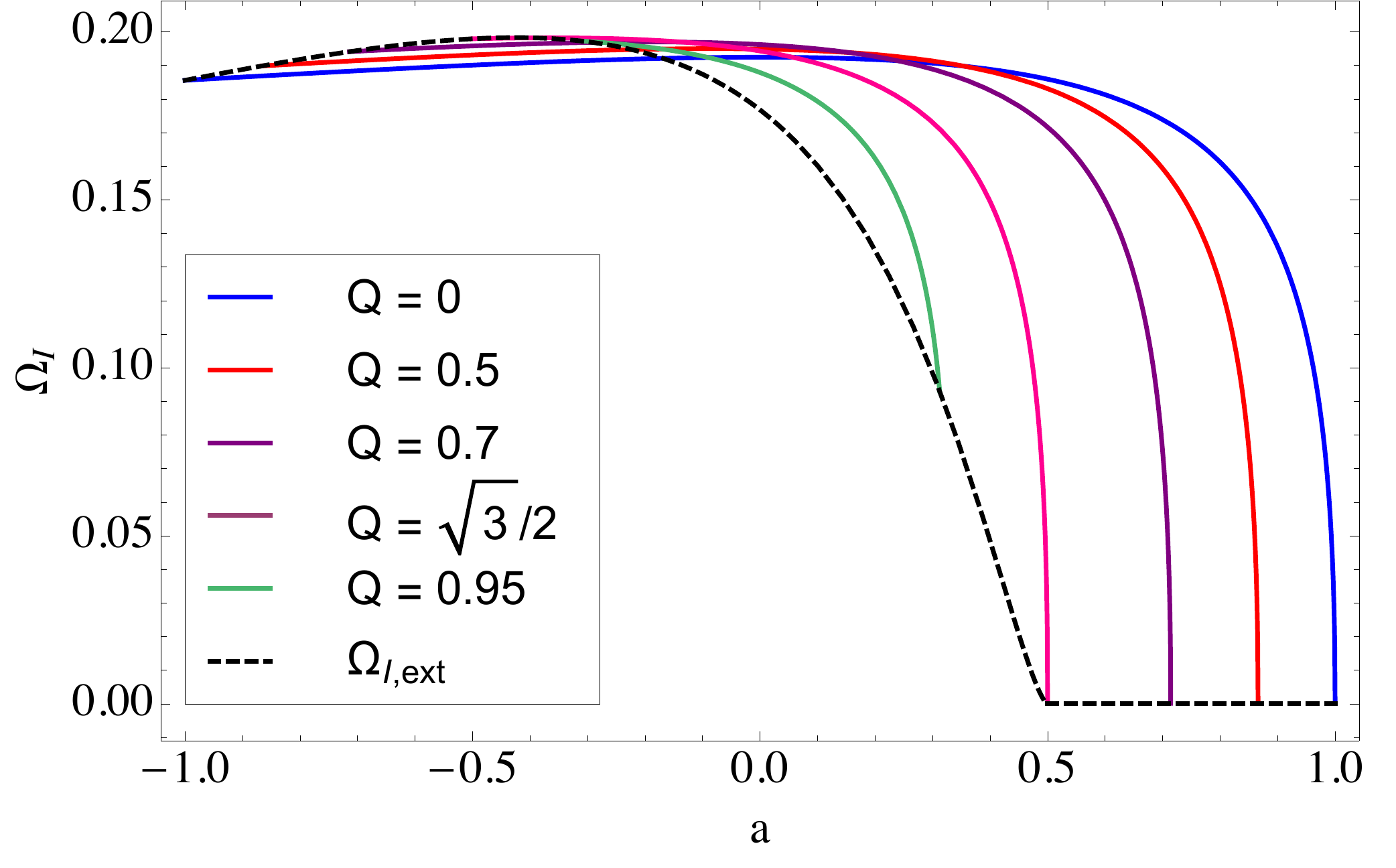}
\caption{{\it Left panel}: The scaled WKB frequency $\Omega_R$ as $a$ varies for fixed values of $Q$, in the case $\mu = 1$. From the far right to left the curves are for $Q=0$ to $Q = 0.95$. Also plotted are the two extremal limits, $\Omega_H$ and $\Omega_{\rm peak}$ [Eq.~\eqref{eq:WKBFreqOuterPeak}].
{\it Right panel}: The scaled WKB decay rate $\Omega_I$ as $a$ varies for the same cases as the top panel. From far right left the curves again are for $Q = 0$ to  
$Q = 0.95$. Also plotted is the extremal prediction $\Omega_{I, {\rm ext}}$ [$0$ for $\mu>\mu_c$ and otherwise given by Eq.~\eqref{eq:DMExDecay}].
}
\label{fig:WKBFreqFixQ}
\end{figure*}

The WKB analysis of the Dudley-Finley equation in the Kerr-Newman spacetime is a straightforward extension of the methods discussed in~\cite{Schutz:1985zz,IyerWill1987} and later extended for generic orbits in Kerr~\cite{Yang:2012he}. 
The result of this analysis gives Eqs.~\eqref{eq:WKBSolve} and~\eqref{eq:WKBOmegaI} for the WKB frequency and decay rates, using the leading order parts of the radial potential for the DF equation.
We present the relevant equations here in full, and specialize to the near-extremal case in Sec.~\ref{sec:WKBanalysis}. For convenience in this section we set $M=1$.
Solving both of the equations of \eqref{eq:WKBSolve} while eliminating $\lambda$ gives our useble formula for $\Omega_R$,
\begin{align}
\label{eq:WKBOmegaR}
\Omega_R & = \frac{\mu a (r - 1)}{(r^2 +a^2)(r -1) - 2 r \Delta}\,,
\end{align}
which must be evaluated at the peak $r_0$ to give a consistent solution to the WKB equations. The above is only an implicit equation for $\Omega_R$ unless we can determine $r_0$ independently of $\Omega_R$. 
Equation~\eqref{eq:WKBOmegaR} shows that if $r_0$ approaches the horizon, $\Omega_R$ approaches the horizon frequency $\Omega_H$. 
This is in agreement with the situation in Kerr, and is only modified by the fact that $\Omega_H$ depends on both $a$ and $Q$.

Using both conditions of Eqs.~\eqref{eq:WKBSolve} together with the approximate analytic expression \eqref{eq:AppxA} for $\alpha(\mu,a,Q)$ lets us eliminate $\Omega_R$, yielding a sixth order polynomial equation, 
\begin{align}
\label{eq:WKBPoly}
2 r^2[r(r-3)+2Q^2]^2 + 4 r\left(r[r^2(1-\mu^2) - 2r -3 (1-\mu^2)]+2Q^2(1-\mu^2 +r) \right)a^2
\notag \\ 
+(1-\mu^2)[r^2(2-\mu^2) + 2r(2+\mu^2) + 2-\mu^2]a^4 =0 \,.
\end{align}
The outermost root of this polynomial gives the position of the peak $r_0$, and when Eq.~\eqref{eq:WKBOmegaR} is evaluated at $r_0$ we attain a self-consistent solution to the equations. 
Note that in the $a \to 0$ case, both the numerator and denominator of Eq.~\eqref{eq:WKBOmegaR} vanish. A better behaved expression in this limit can be found by using the polynomial~\eqref{eq:WKBPoly} to eliminate the vanishing denominator; after some simplification we find
\begin{align}
\label{eq:WKBOmegaRSecond}
\Omega_R = \frac{\sqrt{2}(r - M)}{\sqrt{4r(r^3 - 3 r +2 Q^2) + a^2[(r^2+M)(3-\mu^2)+2r(1+\mu^2)]}}\,.
\end{align}
This equation also behaves correctly in the case $\mu = 0$, for which additional, closed form analytic expressions can be derived for $r_0$ and the frequencies (see e.g~\cite{Teo2003,Dolan2010,Yang:2012he} for the Kerr case). Meanwhile, it is poorly behaved when $r_0$ approaches the horizon in the extremal case, and so is not used outside of this Appendix. 

The WKB analysis gives an equation for $\Omega_I = \omega_I/(n+1/2)$, Eq.~\eqref{eq:WKBOmegaI}.
Using some algebraic tricks that rely on the conditions in Eq.~\eqref{eq:WKBSolve}, we get
\begin{align}
\Omega_I = \left. \Delta\frac{\sqrt{ 4(6 \Omega_R^2 r^2 -1) + 2 a^2 \Omega_R^2(3-\mu^2) }}{2 \Omega_R(r^2+a^2)^2 - 2\mu a (r^2 +a^2) -a \Delta [a \Omega_R (1+\mu^2) - 2 \mu] } \right|_{r_0} \,.
\end{align}
We see immediately that if $r_0$ goes to the horizon in the extremal limit, the WKB analysis predicts a vanishing $\omega_I$, which indicates the existence of ZDMs. 

In Fig.~\ref{fig:WKBFreqFixQ} we plot some representative values of $\Omega_R$ and $\Omega_I$ at fixed charge $Q$ and maximum inclination parameter $\mu = 1$, for values of $a$ varying between each extremal case. For positive values of $a$, the WKB modes correspond to corotating equatorial photon orbits, while for negative values of $a$ they correspond to counterrotating equatorial orbits.

\end{widetext}

\bibliography{KNArxiv2}

\end{document}